\documentclass[12pt]{article}
\pdfoutput=1

\usepackage{jheppub}
\usepackage{amssymb,amsfonts}
\usepackage{epsfig}
\usepackage{tikz}
\usepackage{enumerate}
\usepackage{mathrsfs}
\usepackage{graphicx}
\usepackage{latexsym}
\usepackage{enumerate}

\newcommand{\be}{\begin{equation}}
\newcommand{\ee}{\end{equation}}
\newcommand{\bea}{\begin{eqnarray}}
\newcommand{\eea}{\end{eqnarray}}

\newcommand{\CH}{{\cal H}}

\newcommand{\Ap}{{A'}}
\newcommand{\Bp}{{B'}}
\newcommand{\Cp}{{C'}}
\newcommand{\Dp}{{D'}}

\newcommand{\qd}{{\mathcal{D}}}

\newcommand{\lr}{\left (}
\newcommand{\rr}{\right )}
\newcommand{\ls}{\left [}
\newcommand{\rs}{\right ]}

\newcommand{\p}{\partial}

\renewcommand{\bar}[1]{\overline{#1}}

\makeatletter
\renewcommand{\@seccntformat}[1]{\csname the#1\endcsname.\,\,}
\makeatother

\let \savenumberline \numberline
\def \numberline#1{\savenumberline{#1.}}

\makeatletter
\def\@fpheader{\relax}
\makeatother

\allowdisplaybreaks

\title{\ \vspace{1.6cm} \\
{\scalebox{0.85}{Nonrelativistic String Theory in Background Fields}}}
\author{Jaume Gomis${}^{\, a}$, Jihwan Oh${}^{\, a, b}$, and Ziqi Yan${}^{\, a}$}
\emailAdd{jgomis@perimeterinstitute.ca}
\emailAdd{jihwanoh@berkeley.edu}
\emailAdd{\,\,\,\,\,\,\,\,\,\,\,\,\,\,\,\,\,\,\,\,\,zyan@perimeterinstitute.ca}
\affiliation{
${}^a$Perimeter Institute for Theoretical Physics\\
31 Caroline St N, Waterloo, ON N2L 6B9, Canada\medskip\\
${}^b$Berkeley Center for Theoretical Physics and Department of Physics\\
University of California, Berkeley, CA 94720-7300, USA}
\abstract{Nonrelativistic string theory is a unitary, ultraviolet finite quantum  gravity theory with a nonrelativistic string spectrum. 
The vertex operators of the worldsheet theory determine the spacetime geometry  of  nonrelativistic string theory,   known as the string Newton-Cartan geometry.  We compute the Weyl anomaly of the nonrelativistic string worldsheet sigma model describing  strings propagating in a  string Newton-Cartan geometry, Kalb-Ramond and dilaton   background. 
We derive the equations of motion that dictate the  backgrounds on which nonrelativistic string theory can be consistently defined quantum mechanically. The equations of motion we find from our study of the conformal anomaly of the worldsheet theory are to nonrelativistic string theory what the (super)gravity  equations of motion are to relativistic string theory.}

\begin{document}

\maketitle

\section{Introduction}

The relativistic field equations describing the propagation of massless particles --  Einstein's, Yang-Mills's and Dirac's equation -- beautifully emerge in (relativistic) string theory by demanding consistency of the  quantum field theory  (QFT) living on the string worldsheet.  The spacetime fields, including the spacetime metric, $B$-field and dilaton, are classically marginal couplings   of the two-dimensional quantum field theory on the worldsheet. The fundamental principle determining the spacetime equations of motion   is   Weyl symmetry of the string worldsheet theory, whereby   the spacetime background fields  are tuned to define a two-dimensional conformal field theory. It is the absence of a Weyl anomaly on the worldsheet that  determines the   spacetime dynamics in string theory.

Nonrelativistic string theory in flat spacetime was formulated in \cite{Gomis:2000bd} as a two-dimensional relativistic  QFT on the worldsheet with a nonrelativistic global symmetry. This string theory is unitary, ultraviolet complete and has a spectrum of string excitations with a (string)-Galilean invariant dispersion relation,   and a characteristic string perturbation theory \cite{Gomis:2000bd}.\,\footnote{The nonrelativistic spectrum was first obtained by taking a limit in \cite{Klebanov:2000pp} (see also \cite{Danielsson:2000gi}).} Building on a series of works \cite{Gomis:2005pg, nrGalilei, stringyNC, NCbranes, stringyNClimit}, it was shown in \cite{Bergshoeff:2018yvt} that the appropriate spacetime geometry that nonrelativistic string theory couples to is the so-called string Newton-Cartan geometry.\,\footnote{For other recent work on nonrelativistic strings, see \cite{Batlle:2016iel, Gomis:2016zur, Batlle:2017cfa, HHO, Kluson, Kluson:2018grx, Harmark:2018cdl, Kluson:2018vfd, Kluson:2019ifd, Roychowdhury:2019qmp, Roychowdhury:2019vzh}. In \cite{HHO, Kluson, Harmark:2018cdl}, for zero torsion, a specific truncation of the string Newton-Cartan gravity (with zero $B$-field and dilaton) in the target space was considered, which leads to Newton-Cartan gravity in one dimension lower, supplemented with an extra worldsheet scalar parametrizing the spatial foliation direction. A more thorough examination of this relation is later explored in \cite{Harmark:2019upf}.} In string Newton-Cartan geometry, there is a two-dimensional foliation that naturally splits the target space into a longitudinal and transverse sector. These two sectors are related to each other by (string)-Galilean boosts, which are nontrivially realized on the worldsheet fields of \cite{Gomis:2000bd}. In particular, there is no Riemannian, Lorentzian metric in the target space. Nonrelativistic string theory provides a quantization of string Newton-Cartan gravity  in as much as    relativistic string theory provides a quantization of Einstein's gravity. 

 String theory in a curved background is constructed by deforming the worldsheet theory in flat spacetime by appropriate vertex operators. The fact that the worldsheet fields  and vertex operators of relativistic string theory are different than those of nonrelativistic string theory is what endows these string theories with physically distinct spacetime geometries: a Lorentzian Riemannian metric and  a string Newton-Cartan structure, respectively. Relativistic string theory deformed by massless vertex operators  which represent a condensate of the graviton, $B$-field and dilaton defines the relativistic string sigma model. Nonrelativistic string theory deformed by suitable vertex operators gives rise to strings propagating in a string Newton-Cartan geometry, $B$-field and dilaton background. This deformation defines the nonlinear sigma model for nonrelativistic string theory \cite{Bergshoeff:2018yvt}.\,\footnote{The BRST cohomology of physical states of  nonrelativistic string theory requires that the vertex operators   carry nonzero winding number. One could in principle deform both relativistic or nonrelativistic string theory using vertex operators with winding number. This would be akin to defining the   string field theory geometry. This would be interesting to investigate.}

In this paper, we derive the equations of motion that the string Newton-Cartan geometry, $B$-field and dilaton must obey for nonrelativistic string theory to be quantum mechanically consistent. Our main result is given in (\ref{eq:resultbf}). These equations follow by demanding that nonrelativistic string theory in   nontrivial   backgrounds fields is   Weyl invariant  quantum mechanically, that is by imposing that  the beta-functions for the worldsheet couplings of nonrelativistic string theory vanish. We determine  the background equations of motion  by   computing the linearized  equations  that follow from Weyl invariance 
\begin{equation}
T^\alpha_{~\alpha}=0\, 
\end{equation}
of the vertex operators around flat space, where $T_{\alpha\beta}$ is the energy-momentum tensor on the worldsheet, and then going to higher orders in the background.
This procedure is rather straightforward and makes the origin of the spacetime background equations of motion  physically transparent.  These equations are to nonrelativistic string theory what the supergravity equations of motion are to relativistic string theory. 
The vanishing of the  Weyl anomaly of the worldsheet  QFT defining nonrelativistic string theory determines  the consistent nonrelativistic string backgrounds.

The paper is organized as follows: In \S\ref{sec:nonrelflat} we recall nonrelativistic string theory in flat spacetime. We   define the renormalization of any given operator on a curved worldsheet by generalizing the normal ordering of composite operators in the conformal gauge. Then, we study the tachyonic and first excited vertex operators and study their quantum Weyl transformations. In \S\ref{sec:nstcurved} we study nonrelativistic string theory in curved string Newton-Cartan background with a Kalb-Ramond and dilaton field. After a brief review of the basic ingredients in \S\ref{sec:ingrecurve}, we relate in \S\ref{sec:lbWt} the Weyl transformations of the vertex operators derived in \S\ref{sec:fevo} to beta-functions for the background fields. In \S\ref{sec:nonlinear}, we extend the linearized beta-functions to include higher order terms and present the full beta-functions in \eqref{eq:resultbf}, at the leading $\alpha'$-order.
In \S\ref{sec:conclusions} we conclude the paper and discuss a few future directions.

\vspace{2mm}

 {\it Note Added:}  In the final stages of  this work, we  heard  from Domingo Gallegos, Umut Gursoy and Natale Zinnato of  their study of   Weyl invariance of sigma model in a torsional Newton-Cartan background \cite{Gallegos:2019icg}.

\section{Nonrelativistic String Theory in Flat Spacetime} \label{sec:nonrelflat}

\subsection{The Classical Action} \label{sec:classical}

Nonrelativistic string theory is described by a sigma model defined on a Riemann surface $\Sigma$\,, parametrized by $\sigma^\alpha$\,, $\alpha = 1, 2$\,. The worldsheet metric is denoted by $h_{\alpha\beta}$\,. We introduce the Zweibein field $e_\alpha{}^a$\,, $a = 1, 2$ on $\Sigma$ such that
\be
	h_{\alpha\beta} = e_\alpha{}^a e_\beta{}^b \delta_{ab}\,.
\ee
The worldsheet fields of nonrelativistic string theory consist of worldsheet scalars parametrizing the spacetime coordinates $x^\mu$\,, $\mu = 0, 1, \cdots, d-1$ and two additional   worldsheet fields, which we denote by $\lambda$ and $\overline{\lambda}$\,. We take the decomposition $x^\mu = (x^A, x^\Ap)$\,, with $A = 0,1$ and $\Ap = 2, \cdots, d-1$\,. In terms of the light-cone coordinates,
\be
	X \equiv x^0 + x^1\,, 
		\qquad
	\overline{X} = x^0 - x^1\,,
\ee
the sigma model action is given by \cite{Gomis:2000bd,Bergshoeff:2018yvt} 
\be \label{eq:actionflat}
	S_\text{f} = \frac{1}{4\pi\alpha'} \int d^2 \sigma \sqrt{h} \bigl( \, h^{\alpha\beta} \p_\alpha x^\Ap \p_\beta x^\Bp \delta_{\Ap\Bp} + \lambda \, \overline{\qd} X + \overline{\lambda} \, \qd \overline{X} \, \bigr)\,,
\ee
where $\alpha'$ is the Regge slope, $h = \det h_{\alpha\beta}$ and $h^{\alpha\beta}$ is the inverse of $h_{\alpha\beta}$\,. We defined 
\be
	\qd \equiv \frac{i}{\sqrt{h}} \, \epsilon^{\alpha\beta} \overline{e}_\alpha \nabla_\beta\,, 
		\qquad%
	\overline{\qd} \equiv \frac{i}{\sqrt{h}} \, \epsilon^{\alpha\beta} e_\alpha \nabla_\beta\,,
\ee
where $\nabla_\alpha$ is the covariant derivative compatible with $h_{\alpha\beta}$\,.
%
%
The Levi-Civita symbol $\epsilon^{\alpha\beta}$ satisfies $\epsilon^{12} = - \epsilon^{21} = 1$\,. Moreover, we introduced the (Euclideanized) light-cone coordinates for the flat index $a$ on the worldsheet tangent space,
\be
	e_\alpha = e_\alpha{}^1 + i e_\alpha{}^2\,,
		\qquad
	\overline{e}_\alpha = - e_\alpha{}^1 + i e_\alpha{}^2\,.
\ee

In the conformal gauge, we take
\be \label{eq:cg}
	e_1{}^1 = e_2{}^2 =1\,,
		\qquad
	e_1{}^2 = e_2{}^1 = 0\,,
\ee
and the string action \eqref{eq:actionflat} becomes
\be \label{eq:actionflatcg}
	S_0 = \frac{1}{4\pi\alpha'} \int d^2 z \, \bigl( \, 2 \p_z x^\Ap \p_{\overline{z}} x^\Bp \delta_{\Ap\Bp} + \lambda \, \p_{\overline{z}} X + \overline{\lambda} \, \p_z \overline{X} \, \bigr)\,,
\ee
where we introduced the complex coordinates,
\be \label{eq:zzbar}
	z = \sigma^1 + i \sigma^2\,,
		\qquad
	\overline{z} = \sigma^1 - i \sigma^2\,.
\ee
The equations of motion are
\be \label{eq:eom}
	\p_z \p_{\overline{z}} x^\Ap = \p_{\overline{z}} \lambda = \p_{\overline{z}} X = \p_z \overline{\lambda} = \p_z \overline{X} = 0\,,
\ee
which imply that $\lambda$ and $X$ are holomorphic and $\overline{\lambda}$ and $\overline{X}$ are anti-holomorphic.

We now show  that $\lambda$ and $\overline{\lambda}$ are worldsheet $(1,0)$ and $(0,1)$ forms, respectively, after fixing the conformal gauge. 
First, consider the infinitesimal local transformations of worldsheet fields in the   action $S_\text{f}$ in \eqref{eq:actionflat},
\begin{subequations}
\begin{align}
	\delta e_\alpha & = \xi^\beta \p_\beta e_\alpha + e_\beta \p_\alpha \xi^\beta + \bigl( \omega  + \gamma \, \bigr) e_\alpha\,,
		&
	\delta \lambda & = \xi^\beta \p_\beta \lambda - \bigl( \omega + \gamma \, \bigr) \lambda\,,\\[2pt]
	\delta \overline{e}_\alpha & = \xi^\beta \p_\beta \overline{e}_\alpha + \overline{e}_\beta \p_\alpha \xi^\beta + \bigl( \omega - \gamma \, \bigr) \overline{e}_\alpha\,, 
		& %
	\delta \overline{\lambda} & = \xi^\beta \p_\beta \overline{\lambda} - \bigl( \omega - \gamma \, \bigr) \overline{\lambda}\,, \\[2pt]
		& &
	\delta x^\mu & = \xi^\beta \p_\beta x^\mu\,, 
\end{align}
\end{subequations}
where $\xi_\alpha$ parametrizes   worldsheet diffeomorphisms, and $\omega$ and $\gamma$ parametrize the Weyl transformation and the Lorentz boost, respectively.
We adopt the conformal gauge $e_\alpha{}^a = \delta_\alpha^a$\,. Then, $\delta e_\alpha = \delta \overline{e}_\alpha = 0$ implies that
\be
	\omega + \gamma = - \p_z \xi\,,
		\qquad
	\omega - \gamma = - \p_{\overline{z}} \, \overline{\xi}\,,
		\qquad
	{\p}_{\overline{z}} \xi = \p_z \overline{\xi} = 0\,.
\ee
Using the equations of motion
\be
	{\p}_{\overline{z}} {\lambda} = \p_z \overline{\lambda} = {\p}_{\overline{z}} X = \p_z \overline{X} = 0\,,
\ee
we derive 
\be
	\delta x^\Ap = \xi \, \p_z x^\Ap + \overline{\xi} \, {\p}_{\overline{z}} x^\Ap\,,
		\qquad%
	\delta X = \xi \, \p X\,,
		\qquad%
	\delta \overline{X} = \overline{\xi} \, \p_{\overline{z}} \overline{X}\,,
\ee
and
\be 
	\delta \lambda = \xi \, \p_z \lambda + \lambda \, \p_z \xi\,,
		\qquad%
	\delta \overline{\lambda} = \overline{\xi} \, {\p}_{\overline{z}} \overline{\lambda} + \overline{\lambda} \, \p_{\overline{z}} \overline{\xi}\,.
\ee
This concludes the proof that $\lambda$ and $\overline{\lambda}$ are worldsheet one-forms.

The worldsheet action (\ref{eq:actionflatcg})  was shown in \cite{Gomis:2000bd} to lead to a unitary, ultraviolet finite string theory with a well-defined perturbative expansion. 

\subsection{Renormalized Operators and Weyl Transformations} \label{sec:renorm}

In this subsection, we define how a given operator constructed out of the worldsheet fields in nonrelativistic string theory is renormalized and study the Weyl transformations of these renormalized operators. We will follow closely the techniques and conventions in \cite{Polchinski:1998rq}.

The nontrivial operator product expansions (OPEs) among the various worldsheet fields in the nonrelativistic string action \eqref{eq:actionflatcg} are given as follows:
\begin{subequations} \label{eq:normalordering}
\begin{align}
	x^\Ap (z, \overline{z}) \, x^\Bp (w, \overline{w}) & \sim - \tfrac{1}{2} \alpha' \delta^{\Ap\Bp} \ln | z - w |^2\,, \\[2pt]
	\lambda (z, \overline{z}) \, X(w, \overline{w}) & \sim - \frac{2\alpha'}{z - w}\,, \\[2pt]
	\overline{\lambda} (z, \overline{z}) \, \overline{X}(w, \overline{w}) & \sim - \frac{2\alpha'}{\overline{z} - \overline{w}}\,.		
\end{align}
\end{subequations}
Then, the normal ordering of a given operator $\mathcal{O}$ can be compactly written as
\be \label{eq:normalorderingO}
	: \! \mathcal{O} \! : \, = \exp \ls \frac{\alpha'}{4} \int d^2 z \, d^2 w \ln |z - w|^2 \, \widehat{P} (z, \overline{z}; w, \overline{w}) \rs \mathcal{O}\,,
\ee
where
\begin{align} 
	\widehat{P}(z, \overline{z}; w, \overline{w}) & = \frac{\delta}{\delta x^{\Ap} (z\,, \overline{z})} \frac{\delta}{\delta x^{\Ap} (w\,, \overline{w})} \notag \\[4pt]
		& \quad - 8 \biggl[ \, \p_{z} \frac{\delta}{\delta \lambda(z\,, \overline{z})} \frac{\delta}{\delta X(w\,, \overline{w})} + \p_{\overline{z}} \frac{\delta}{\delta \bar\lambda(z\,, \overline{z})} \frac{\delta}{\delta \bar X(w\,, \overline{w})} \, \biggr]\,.
\end{align}
The normal ordering defined in \eqref{eq:normalorderingO} guarantees that the equations of motion in \eqref{eq:eom} hold as operator equations.

Further note that the path integral of a total derivative is identically zero. For example, for $S_0$ in \eqref{eq:actionflatcg}, from
\be
	0 = \int \mathscr{D} \lambda \, \mathscr{D} \bar\lambda \, \mathscr{D} x^\mu \frac{\delta}{\delta x^\Ap (z, \overline{z})} \exp(-S_0)\,, 
\ee
we derive the following equation of motion that holds as an operator equation:
\be
	\langle \p_z \p_{\overline{z}} x^\Ap \rangle = 0\,.
\ee		
Similarly, we have
\be
	\langle \p_z \bar\lambda \rangle = \langle \p_{\overline{z}} \lambda \rangle = \langle \p_{\overline{z}} X \rangle = \langle \p_{z} \overline{X} \rangle = 0\,.
\ee
The normal ordering defined in \eqref{eq:normalorderingO} is compatible with the above Ehrenfest theorems, i.e. the following operator equations hold:
\be \label{eq:operatoreom}
	: \! \p_z \p_{\overline{z}} x^\Ap \!\!\! : \, = \, : \! \p_z \bar\lambda \! : \, = \, : \! \p_{\overline{z}} \lambda \! : \, = \, : \! \p_{\overline{z}} X \! : \, = \, : \! \p_{z} \overline{X} \! : \, = \,  0\,.
\ee

With the normal ordering defined in \eqref{eq:normalorderingO}, we proceed to the BRST quantization of \eqref{eq:actionflatcg}. The gauge fixed action is
\be 
	S_\text{BRST} = \frac{1}{4\pi\alpha'} \int d^2 z \, \Bigl[ \, 2 \p_z x^\Ap \p_{\overline{z}} x^\Bp \delta_{\Ap\Bp} + \lambda \, \p_{\overline{z}} X + \overline{\lambda} \, \p_z \overline{X}  + 2 \alpha' \bigl( b \, \p_{\overline{z}} c + \overline{b} \, \p_z \overline{c} \bigr) \, \Bigr]\,. 	
\ee
Note that the $bc$ ghost system is exactly the same as in relativistic string theory. The BRST transformations are\,\footnote{We only present the transformations for the unbarred fields; the barred ones follow accordingly.}
\begin{subequations}
\begin{align}
	\delta_B x^\Ap & = i \epsilon (c \, \p_z + \overline{c} \, {\p}_{\overline{z}}) x^\Ap\,, 
		&%
	\delta_B X & = i \epsilon \, c \, \p_z X\,, 
		&%
	\delta \lambda & = i \epsilon \bigl( c \, \p_z \lambda + \lambda \, \p_z c \bigr)\,, \\
	\delta_B b & = i \epsilon \bigl( T^\text{m} + T^\text{g} \bigr)\,, 
		&%
	\delta_B c & = i \epsilon \, c \, \p_z c\,,
\end{align}
\end{subequations}
where
\begin{align} \label{eq:TmTg}
	T^\text{m} & = - \frac{1}{2 \alpha'} \bigl( 2 \! : \! \p_z x^\Ap \p_z x^\Ap \!\! : + : \! \lambda \, \p_z X \! : \bigr)\,, 
		&%
	T^\text{g} & = \, :\! \p b \, c \!: - 2 \, \p \! : \! b c \! :\,.
\end{align}
The BRST current is in form the same as in relativistic string theory,
\be
	J_\text{BRST} = c \, T^\text{m} + :\! b c \, \p c \!: + \frac{3}{2} \p^2 c\,,
\ee
but with the matter stress energy tensor given by $T^\text{m}$ in \eqref{eq:TmTg}. Requiring that the associated BRST operator be nilpotent constrains the matter central charge $c_\text{m}$ to be 26. Furthermore, note that $c_\text{m}$ receives a contribution $c_\text{trans.} = d-2$ from the $d-2$ transverse fields $x^\Ap$ and a contribution $c_\text{long.} = 2$ from the longitudinal commuting $\beta\gamma$ system. In total, we have $c_\text{m} = c_\text{trans.} + c_\text{long.} = d$\,. Therefore, the nilpotence of $Q_\text{BRST}$ requires the critical dimension to be 26 \cite{Gomis:2000bd}, akin to relativistic string theory.

In order to define    normal ordering   on a curved worldsheet in a diffeomorphism invariant way, we need to invoke  the geodesic distance $d(\sigma_1, \sigma_2)$ between two points $\sigma_1$ and $\sigma_2$ on $\Sigma$\, . The   renormalized operator $[\mathcal{O}]_\text{r}$  for an operator $\mathcal{O}$  is given by
\begin{align} \label{eq:renormalizationO}
	[\mathcal{O}]_\text{r} \, & = \exp \ls \frac{\alpha'}{4} \int d^2 \sigma_1 \, d^2 \sigma_2 \ln \bigl[ d (\sigma_1, \sigma_2) \bigr]^2 \, \widehat{P} (\sigma_1, \sigma_2) \rs \mathcal{O}\,, 
\end{align}
where
\begin{align} 
	\widehat{P}(\sigma_1, \sigma_2) & = \frac{\delta}{\delta x^{\Ap} (\sigma_1)} \frac{\delta}{\delta x^{\Ap} (\sigma_2)} - 4 \biggl[ \, \qd_{\sigma_1} \frac{\delta}{\delta \lambda(\sigma_1)} \frac{\delta}{\delta X(\sigma_2)} + \overline{\qd}_{\sigma_1} \frac{\delta}{\delta \bar\lambda(\sigma_1)} \frac{\delta}{\delta \bar X(\sigma_2)} \, \biggr]\,.
\end{align}
In the flat worldsheet limit, \eqref{eq:renormalizationO} gives back \eqref{eq:normalorderingO}.

While the renormalized operator (\ref{eq:renormalizationO}) is diffeomorphism invariant, it is not necessarily Weyl invariant; the renormalization of the operator may lead to a Weyl anomaly. Under the Weyl transformations generated by
\be \label{eq:Weyldef}
	\delta_W e_\alpha{}^a = \delta \omega \, e_\alpha{}^a\,,
		\qquad%
	\delta_W \lambda = - \delta \omega \, \lambda\,,
		\qquad
	\delta_W \overline{\lambda} = - \delta \omega \, \bar\lambda\,,
\ee
the renormalized operator $\bigl[ \mathcal{O} \bigr]_\text{r}$ transforms as\,\footnote{It is easy to show that $\widehat{P}(\sigma_1, \sigma_2)$ in  (\ref{eq:renormalizationO})  is Weyl invariant.}
\be \label{eq:WeylO}
	\delta_W \bigl[ \mathcal{O} \bigr]_\text{r} = [\delta_W \mathcal{O}]_\text{r} + \exp \ls \frac{\alpha'}{2} \int d^2 \sigma_1 \, d^2 \sigma_2 \Bigl( \delta_W \ln d (\sigma_1, \sigma_2) \Bigr) \widehat{P} (\sigma_1, \sigma_2) \rs \mathcal{O}\,.
\ee
It is useful to note the coincidence limit of Weyl transformations of $\ln d(\sigma, \sigma')$ \cite{Polchinski:1998rq}:
\begin{subequations} \label{eq:lndtransformations}
\begin{align}
	\delta_W \ln d(\sigma, \sigma') \big|_{\sigma' = \sigma} & = \delta \omega (\sigma)\,, \\[2pt]
	\p_\alpha \delta_W \ln d(\sigma, \sigma') \big|_{\sigma' = \sigma} & = \tfrac{1}{2} \p_\alpha \delta \omega (\sigma)\,, \\[2pt] 
	\p_\alpha \p'_\beta \delta_W \ln d(\sigma, \sigma') \big|_{\sigma' = \sigma} & = \tfrac{1}{2} (\gamma+1) \, \nabla_\alpha \p_\beta \delta \omega (\sigma)\,, \\[2pt]
	\nabla_\alpha \p_\beta \delta_W  \ln d(\sigma, \sigma') \big|_{\sigma' = \sigma} & = - \tfrac{1}{2} \gamma \, \nabla_\alpha \p_\beta \delta \omega (\sigma)\,.
\end{align}
\end{subequations}
The constant  $\gamma$  parametrizes  the  renormalization scheme. 

\subsection{Tachyon Vertex Operator}

In nonrelativistic string theory, the string spectrum is empty unless one compactifies the longitudinal spatial $x^1$ direction on a circle of radius $r$\,, and all states in the Hilbert space necessarily have nonzero windings. Scattering amplitudes of tachyonic winding states in nonrelativistic string theory were computed in \cite{Gomis:2000bd}. Consider the vertex operator for a closed string tachyon with winding number $w$\,,\,\footnote{The expression of the vertex operator involving $\lambda$ can be given an explicit representation in terms of  its mode expansion.}
\be \label{eq:top}
	V_\text{t} = g_\text{t} : e^{i\ell} :\,, 
		\qquad%
	\ell \equiv k_\Ap x^\Ap + p \, X + \overline{p} \, \overline{X} + q \int^z \lambda + \overline{q} \int^{\overline{z}} \overline{\lambda}\,,
\ee
where $k_\Ap$ is the transverse momentum and $p$\,, $\overline{p}$ are the longitudinal momenta, with
\be \label{eq:pbarp}
	p = \frac{1}{2} (p_0 + p_1)\,,
		\qquad%
	\overline{p} = \frac{1}{2} (p_0 - p_1)\,.
\ee
The ``momenta" $q$ and $\overline{q}$ encode the windings,
\be
	q = \overline{q} = - \frac{wr}{4\alpha'}\,. 
\ee
The conformal weight of $V_\text{t}$ is 
\be \label{eq:cwVt}
	h = \frac{1}{4} \bigl( \alpha' k^2 - 2 wr \, p \bigr)\,,
		\qquad
	\overline{h} = \frac{1}{4} \bigl( \alpha' k^2 - 2 wr \, \overline{p} \bigr)\,,
\ee
where $k^2 \equiv k_\Ap k^\Ap$\,.

Integrated over a curved worldsheet, $V_\text{t}$ gives rise to
\be \label{eq:Vt}
	\mathcal{V}_\text{t} = g_\text{t} \int d^2 \sigma \sqrt{h} \, \bigl[ e^{i  \ell} \bigr]_\text{r}\,,
\ee
where
\be
	\ell \equiv k_\Ap x^\Ap + \, p \, X + \, \overline{p} \, \overline{X} + \, q \int^\sigma \lambda \, i \epsilon^\alpha{}_\beta \, e_\alpha \, d\sigma^\beta + \, \overline{q} \int^\sigma  \overline{\lambda} \, i \epsilon^\alpha{}_\beta \, \overline{e}_\alpha \, d\sigma^\beta\,.
\ee
Here, $\epsilon^\alpha{}_\beta \equiv \epsilon^{\alpha\gamma} h_{\beta\gamma}$\,. The analyticity condition on $\lambda$ becomes
\be
	i \epsilon^{\alpha\beta} e_\alpha{}^1 \, \p_\beta \lambda = \epsilon^{\alpha\beta} e_\alpha{}^2 \, \p_\beta \lambda\,.
\ee
With respect to the Weyl transformations \eqref{eq:Weyldef}, applying \eqref{eq:WeylO} to \eqref{eq:Vt} gives the Weyl transformation of $\mathcal{V}_t$\,,
\be
	\delta_W \mathcal{V}_\text{t} = g_\text{t} \int d^2 \sigma \sqrt{h} \, \delta \omega \, \bigl[ 2 - \tfrac{1}{2} \alpha' k^2 + \tfrac{1}{2} wr \, (p + \overline{p}) \bigr]\,.
\ee
Requiring that $\delta_W \mathcal{V}_\text{t}$ vanish gives rise to the tachyon on-shell condition
\be
	p_0 = \frac{\alpha'k^2 - 4}{w r}\,.
\ee
This is the same as requiring that $V_\text{t}$ be a $(1,1)$-form by setting $h = \overline{h} = 1$ in \eqref{eq:cwVt}, which also sets $p_1 = 0$\,.

In general, the four tachyon amplitude has poles corresponding to intermediate   excited closed string states carrying nonzero windings. However, in the special case in which the winding number is not exchanged among strings, the four tachyon amplitude also gains a contribution from exchanging off-shell states in the zero winding sector. The leading long-range contribution is proportional to $1 / k^2$\,, with $k_\Ap$ the off-shell transverse momentum of the intermediate state, which gives rise to the (string) Newtonian potential after  a Fourier transform. Note that, in the zero winding sector, all on-shell strings have $k_\Ap = 0$\,. Therefore, these zero winding intermediate states become of measure zero as asymptotic states; but they nevertheless mediate an instantaneous gravitational force between winding strings \cite{Gomis:2000bd, Danielsson:2000mu}.  

\subsection{Vertex Operators for String Newton-Cartan Backgrounds} \label{sec:fevo}

In this subsection, we consider the vertex operators corresponding to the zero winding  states that mediate Newtonian forces between winding strings, for which we compute the Weyl transformations. These determine the string Newton-Cartan background fields on which nonrelativistic strings can consistently propagate.\,\footnote{One could extend the analysis to the geometric structure that describes the theory when a condensate of string states with winding number is nontrivial.}

First, we classify the first excited vertex operators that are $(1,1)$-forms as follows:
\be
	: \! \p_z x^\mu \, \p_{\overline{z}} x^\nu \, e^{i\ell} \! :\,,
		\qquad
	: \! \lambda \, \p_{\overline{z}} x^\mu \, e^{i\ell} \! :\,, 
		\qquad
	: \! \bar\lambda \, \p_z x^\mu \, e^{i\ell} \! :\,,
		\qquad
	: \! \lambda \, \bar\lambda\, e^{i\ell} \! :\,.
\ee
On the curved worldsheet, these operators become, respectively,
\be \label{eq:curvedvertexoperator}
	\bigl[ \qd x^\mu \, \overline{\qd} x^\nu \, e^{i\ell} \bigr]_\text{r}\,,
		\qquad
	\bigl[ \lambda \, \overline{\qd} x^\mu \, e^{i\ell} \bigr]_\text{r}\,,
		\qquad
	\bigl[ \bar\lambda \, \qd x^\mu \, e^{i\ell} \bigr]_\text{r}\,,
		\qquad
	\bigl[ \lambda \, \bar\lambda\, e^{i\ell} \bigr]_\text{r}\,.
\ee
Including the contribution from the dilaton and integrating over the worldsheet, we obtain the worldsheet diffeomorphism invariant vertex operator,
\begin{align} \label{eq:fevobefore}
	\mathcal{V} = \frac{g}{\alpha'} \int d^2 \sigma \sqrt{h} \, \Bigl\{ (s_{\mu\nu} + a_{\mu\nu}) \bigl[ \qd x^\mu \, \overline{\qd} x^\nu \, e^{i\ell} \bigr]_\text{r}
		& 	+ t_\mu \bigl[ \lambda \, \overline{\qd} x^\mu e^{i\ell} \bigr]_\text{r}
			+ \overline{t}_\mu \bigl[ \bar\lambda \, \qd x^\mu e^{i\ell} \bigr]_\text{r} \notag \\
		& 	+ \, u \, \bigl[ \lambda \, \bar\lambda\, e^{i\ell} \bigr]_\text{r}
			\hspace{-0.5mm} + \alpha' f R^{(2)} \bigl[ e^{i\ell} \bigr]_\text{r}
		\, \Bigr\}\,.
\end{align}
Here, $R^{(2)}$ is the worldsheet Ricci scalar, $s_{\mu\nu}$ is a symmetric tensor and $a_{\mu\nu}$ is an antisymmetric tensor. Turning on the $\lambda \overline{\lambda}$ operator in \eqref{eq:fevobefore} with a nonzero $u$ generates a deformation towards relativistic string theory.\,\footnote{This can be seen by integrating out $\lambda$ and $\overline{\lambda}$ in the deformed action.} To focus on nonrelativistic string theory (with zero winding), we set $u = 0$\,, which does not receive quantum corrections at linear order or higher in fluctuations as shown later in \eqref{eq:Weyltransformations} and in \S\ref{sec:nonlinear}. This is protected by the string Newton-Cartan symmetry that will be introduced in \S\ref{sec:ingrecurve}.

Next, we set the winding number to zero but allow the transverse momentum $k_\Ap$ to take any nonzero value. Then, we have in $e^{i \ell}$
\be
	\ell = k_{A'} x^{A'} + p \, X + \overline{p} \, \overline{X}\,.
\ee
Setting the winding number to zero brings significant simplifications. Importantly, note that some of the terms in \eqref{eq:fevobefore} are now trivially zero. Starting with the path integral identity,
\be
	0 = \int \mathscr{D} \lambda \, \mathscr{D} \bar\lambda \, \mathscr{D} x^\mu \frac{\delta}{\delta \lambda (z, \overline{z})} \ls \, \exp(- S_0) \, \p_{w} x^\mu(w, \overline{w}) \, e^{i\ell (w, \overline{w})} \, \rs \biggl |_{(w, \overline{w}) \rightarrow (z, \overline{z})}\,, 
\ee
it follows that
\be
	0 =  \bigl\langle \p_{\overline{z}} X \, \p_{z} x^\mu \, e^{i\ell} \bigr\rangle\,,
\ee
which on the curved worldsheet is regularized to be
\be \label{eq:dXdxe}
	\bigl[ \overline{\qd} X \, {\qd} x^\mu e^{i\ell} \bigr]_\text{r} = 0\,.
\ee
Similarly, we have
\be \label{eq:bdXdxe}
	\bigl[ {\qd} \overline{X} \, \overline{\qd} x^\mu e^{i\ell} \bigr]_\text{r} = 0\,.
\ee
There is another more subtle identity: from
\begin{align}
	0 & = \int \mathscr{D} \lambda \, \mathscr{D} \bar\lambda \, \mathscr{D} x^\mu \frac{\delta}{\delta \lambda (z, \overline{z})} \Bigl[ \, \exp(- S_0) \, \lambda (w, \overline{w}) \, e^{i\ell (w, \overline{w})} \, \Bigr] \,, \notag \\
	0 & = \int \mathscr{D} \lambda \, \mathscr{D} \bar\lambda \, \mathscr{D} x^\mu \frac{\delta}{\delta \overline{\lambda} (z, \overline{z})} \Bigl[ \, \exp(- S_0) \, \overline{\lambda} (w, \overline{w}) \, e^{i\ell (w, \overline{w})} \, \Bigr] \,, 
\end{align}
we obtain
\begin{subequations} \label{eq:lambdaXWard}
\begin{align}
	\bigl\langle \lambda(w, \overline{w}) \, \p_{\overline{z}} X(z, \overline{z}) \, e^{i\ell(w, \overline{w})} \bigr\rangle & = \bigl\langle \delta^{(2)} (z-w, \overline{z}-\overline{w}) \, e^{i\ell(w, \overline{w})} \bigr\rangle\,, \\[2pt] 
	\bigl\langle \overline{\lambda}(w, \overline{w}) \, \p_{z} \overline{X}(z, \overline{z}) \, e^{i\ell(w, \overline{w})} \bigr\rangle & = \bigl\langle \delta^{(2)} (z-w, \overline{z}-\overline{w}) \, e^{i\ell(w, \overline{w})} \bigr\rangle\,.
\end{align}
\end{subequations}
In the coincidence limit $(w, \overline{w}) \rightarrow (z, \overline{z})$\,, \eqref{eq:lambdaXWard} gives
\be
	\bigl\langle \lambda(z, \overline{z}) \, \p_{\overline{z}} X(z, \overline{z}) \, e^{i\ell(z, \overline{z})} \bigr\rangle = \bigl\langle \overline{\lambda}(z, \overline{z}) \, \p_{z} \overline{X}(z, \overline{z}) \, e^{i\ell(z, \overline{z})} \bigr\rangle\,.
\ee
On the curved worldsheet, we have
\be \label{eq:addid}
	\bigl[ \lambda \, \overline{\qd} X e^{i\ell} \bigr]_\text{r} = \bigl[ \overline{\lambda} \, {\qd} \overline{X} e^{i\ell} \bigr]_\text{r}\,.
\ee

Finally, taking \eqref{eq:dXdxe}, \eqref{eq:bdXdxe} and \eqref{eq:addid} into account, we find the worldsheet diffeomorphism invariant vertex operator that describes the zero winding intermediate states and the background fields on which nonrelativistic strings propagate,
\begin{align} \label{eq:fevo}
	\mathcal{V} = \frac{g}{\alpha'} \int d^2 \sigma \sqrt{h} \, \Bigl\{ & \bigl( s_{\Ap\Bp} + a_{\Ap\Bp} \bigr) \bigl[{\qd} x^\Ap \overline{{\qd}} x^\Bp e^{i\ell} \bigr]_\text{r} \notag \\
		& + \theta \,\bigl[{\qd}X \, \overline{{\qd} X} e^{i\ell} \bigr]_\text{r}
		+ \theta_\Ap \bigl[{\qd} X \, \overline{{\qd}} x^\Ap e^{i\ell} \bigr]_\text{r} + \overline{\theta}_\Ap \bigl[\overline{{\qd} X} \, {{\qd}} x^\Ap e^{i\ell} \bigr]_\text{r} \notag \\[4pt]
		& + t_\Ap \bigl[\lambda \, \overline{{\qd}} x^\Ap e^{i\ell} \bigr]_\text{r} 
		+ t \, \bigl[\lambda \, \overline{{\qd}} \overline{X} e^{i\ell} \bigr]_\text{r} 
		+ \overline{t}_\Ap \bigl[\overline{\lambda} \, {{\qd}} x^\Ap e^{i\ell} \bigr]_\text{r} 
		+ \overline{t} \, \bigl[\overline{\lambda} \, {{\qd}} {X} e^{i\ell} \bigr]_\text{r} \notag \\[2pt]
		& + t' \bigl[\lambda \, \overline{{\qd}} X e^{i\ell} \bigr]_\text{r}  
		+ \alpha' f R^{(2)} \bigl[ e^{i\ell} \bigr]_\text{r}\, \Bigr\}\,.
\end{align}
Note that now $\ell = k_\Ap x^\Ap + p \, X + \overline{p} \, \overline{X}$ and
\begin{subequations}
\begin{align}
	t' & = \tfrac{1}{2} \bigl( t_0 + t_1 + \overline{t}_0 - \overline{t}_1 \bigr) \,, & 
	\theta_{A\Ap} & = \tfrac{1}{2} \bigl( s_{A\Ap} + \epsilon_A{}^B a_{B\Ap} \bigr)\,, \\
	t & = \tfrac{1}{2} ( t_0 - t_1 )\,, &
	\theta_{\Ap} & \equiv \theta_{0\Ap} + \theta_{1\Ap}\,, \\
	\overline{t} & = \tfrac{1}{2} ( \overline{t}_0 + \overline{t}_1 )\,, &
	\overline{\theta}_{\Ap} & \equiv \theta_{0\Ap} - \theta_{1\Ap}\,, \\
	& &
	 \theta & = - \tfrac{1}{4} \bigl( s_A{}^A - \epsilon^{AB} a_{AB} \bigr)\,.
\end{align}
\end{subequations}
Here, the Levi-Civita symbol $\epsilon_{AB}$ satisfies $\epsilon_{01} = - \epsilon_{10} = 1$\,. The index $A$ is raised (lowered) by $\eta^{AB}$ ($\eta_{AB}$) with
\be
	\eta^{AB} = 
		\begin{pmatrix}
			-1 & \,\, 0 \, \\
			0 & \,\, 1 \,
		\end{pmatrix}\,.
\ee

In \S\ref{sec:renorm}, we derived the operator equations \eqref{eq:operatoreom}. %
One may also insert in the normal orderings an arbitrary local operator 
\be
	: \cdots \mathcal{O} (w, \overline{w}) :
\ee 
at a point $w$ in the worldsheet other than $z$\,, without affecting any of the above equations. If the equations of motion \eqref{eq:operatoreom} are multiplied with another operator at the same point, there are sometimes additional divergences to be regularized. In particular, if this insertion is the tachyon vertex operator $e^{i\ell}$\,, the divergences can be absorbed into the dilaton counterterm. For example, on the curved worldsheet, we have (see also \cite{Polchinski:1998rq})
\begin{subequations} \label{eq:renormeom}
\begin{align} 
	\bigl[ \overline{\qd} \qd x^\Ap e^{i\ell} \bigr]_\text{r} & = \frac{i}{4} \, \alpha' \gamma \, k^\Ap R^{(2)} \, [e^{i\ell}]_\text{r}\,, \\
	\bigl[ \overline{\qd} \lambda \, e^{i\ell} \bigr]_\text{r} & = \frac{i}{2} \, \alpha' \gamma \, p \, R^{(2)} \, [e^{i\ell}]_\text{r}\,, \label{eq:eq:renormeom2}\\
	\bigl[ {\qd} \overline{\lambda} \, e^{i\ell} \bigr]_\text{r} & = \frac{i}{2} \, \alpha' \gamma \, \overline{p} \, R^{(2)} \, [e^{i\ell}]_\text{r}\,.
\end{align}
\end{subequations}
The coefficients on the right hand side of the above equations are parametrized by the renormalization scheme dependent factor $\gamma$ such that the Weyl transformations of \eqref{eq:renormeom} are consistent with \eqref{eq:lndtransformations}.
This $\gamma$ dependence is, however, absent in $\bigl[ \overline{\qd} \qd x^A e^{i\ell} \bigr]_\text{r}$\,. Note that, in the zero winding sector, $\ell$ is independent of $\lambda$ and $\overline{\lambda}$\,. Therefore, 
\be \label{eq:renormeomxA}
	\bigl[ \overline{\qd} \qd x^A e^{i\ell} \bigr]_\text{r} = 0\,.
\ee

Collecting the relations \eqref{eq:WeylO}, \eqref{eq:renormeom} and \eqref{eq:renormeomxA}, it is a lengthy but straightforward exercise to derive the Weyl transformation of $\mathcal{V}$ in \eqref{eq:fevo}. Recall that, to focus on nonrelativistic string theory, we already set $u = 0$ in \eqref{eq:fevobefore}, and the associated operator $\lambda \overline{\lambda}$ is not generated at linear order or higher in quantum fluctuations (also see \S\ref{sec:nonlinear})\,. 
The resulting Weyl transformation is
\begin{align} \label{eq:Weyltransformations}
	\delta_W \mathcal{V} = \frac{g}{2} \int d^2 \sigma & \sqrt{h} \, \Bigl\{ \bigl( \delta s_{\Ap\Bp} + \delta a_{\Ap\Bp} \bigr) \bigl[{\qd} x^\Ap \overline{{\qd}} x^\Bp e^{i\ell} \bigr]_\text{r} \notag \\
		& + \delta \theta \,\, \bigl[{\qd}X \, \overline{{\qd} X} e^{i\ell} \bigr]_\text{r}
		+ \delta \theta_\Ap \bigl[{\qd} X \, \overline{{\qd}} x^\Ap e^{i\ell} \bigr]_\text{r} + \delta \overline{\theta}_\Ap \bigl[\overline{{\qd} X} \, {{\qd}} x^\Ap e^{i\ell} \bigr]_\text{r} \notag \\[4pt]
		& + \delta t_\Ap \bigl[\lambda \, \overline{{\qd}} x^\Ap e^{i\ell} \bigr]_\text{r} 
		+ \delta t \, \bigl[\lambda \, \overline{{\qd}} \overline{X} e^{i\ell} \bigr]_\text{r} 
		+ \delta \overline{t}_\Ap \bigl[\overline{\lambda} \, {{\qd}} x^\Ap e^{i\ell} \bigr]_\text{r} 
		+\delta \overline{t} \, \bigl[\overline{\lambda} \, {{\qd}} {X} e^{i\ell} \bigr]_\text{r} \notag \\[2pt]
		& + \delta t' \bigl[\lambda \, \overline{{\qd}} X e^{i\ell} \bigr]_\text{r} 
		+ \alpha' \delta f R^{(2)} \bigl[ e^{i\ell} \bigr]_\text{r}\, \Bigr\}\,,
\end{align}
where
\begin{subequations} \label{eq:lightconebeta}
\begin{align}
	\delta s_{A'B'} = \Bigl[ \, & k_{A'} k^{C'} s_{B'C'} + k_{B'} k^{C'} s_{A'C'} - k^2 s_{A'B'} - (\gamma+1) k_{A'} k_{B'} s_{C'}{}^{C'} \notag \\[2pt]
		& + \bigl( p \, k_{A'} t_{B'} + \overline{p} \, k_{B'} \overline{t}_{A'} \bigr) + (p \, k_{B'} t_{A'} + \overline{p} \, k_{A'} \overline{t}_{B'}) \notag \\[2pt]
		& - 2 (\gamma+1) k_{A'} k_{B'} t' + 4 \, k_{A'} k_{B'} f \Bigr] \, \delta \omega \,,\\[4pt]
	\delta a_{A'B'} = \Bigl[ \, & \! - k_{A'} k^{C'} a_{B'C'} + k_{B'} k^{C'} a_{A'C'} - k^2 a_{A'B'} \notag \\[2pt]
		& + \bigl( p \, k_{A'} t_{B'} + \overline{p} \, k_{B'} \overline{t}_{A'} \bigr) - (p \, k_{B'} t_{A'} + \overline{p} \, k_{A'} \overline{t}_{B'}) \Bigr]\, \delta \omega\,, \\[10pt] 
	\delta \theta = \Bigl[ \, & \!  - (\gamma + 1) \, p \, \overline{p} \, s_{A'}{}^{A'} - k^2 \theta + \overline{p} \, k^{A'} \theta_{A'} + {p} \, k^{A'} \overline{\theta}_{A'} \notag \\[2pt]
		& + 2 \, (p^2 t + \overline{p}^2 \overline{t}) - 2 (\gamma + 1) \, p \, \overline{p} \, t' + 4 \, p \, \overline{p} \, f \Bigr]\, \delta \omega \,, \\[2pt]
	\delta {\theta}_{A'} = \Bigl[ \, & {p} \, k^{B'} (s_{A'B'} - a_{A'B'}) - (\gamma + 1) \, p \, k_{A'} s_{B'}{}^{B'} + k_{A'} k^{B'} \theta_{B'} - k^2 \, \theta_{A'} \notag \\[2pt]
		& + 2 \, p^2 \, t_{A'} + 2 \, k_{A'} \overline{p} \, \overline{t} - 2 (\gamma+1) \, p \, k_{A'} \, t' + 4 \, p \, k_{A'} f \Bigr] \delta \omega \,, \\[2pt]
	\delta \overline{\theta}_{A'} = \Bigl[ \, & \overline{p} \, k^{B'} (s_{A'B'} + a_{A'B'}) - (\gamma + 1) \, \overline{p} \, k_{A'} s_{B'}{}^{B'} + k_{A'} k^{B'} \overline{\theta}_{B'} - k^2 \, \overline{\theta}_{A'} \notag \\[2pt]
		& + 2 \, \overline{p}^2 \, \overline{t}_{A'} + 2 \, k_{A'} {p} \, {t} - 2 (\gamma+1) \, \overline{p} \, k_{A'} \, t' + 4 \, \overline{p} \, k_{A'} f \Bigr] \, \delta \omega \,,
\end{align}
and
\begin{align}
	\delta t_{A'} & = \bigl( k_{A'} k^{B'} t_{B'} - k^2 t_{A'} \bigr) \, \delta \omega\,, 
		&%
	\delta t & = \bigl( \overline{p} \, k^{A'} t_{A'} - k^2 t \bigr) \, \delta \omega\,, \\[2pt]
	\delta \overline{t}_{A'} & = \bigl( k_{A'} k^{B'} \overline{t}_{B'} - k^2 \overline{t}_{A'} \bigr) \, \delta \omega\,, 
		&%
	\delta \overline{t} & = \bigl( {p} \, k^{A'} \overline{t}_{A'} - k^2 \overline{t} \bigr) \, \delta \omega\,, \\[2pt]
	& & \delta t' & = \Bigl[ k^{A'} ( p \, t_{A'} + \overline{p} \, \overline{t}_{A'} ) - k^2 t' \Bigr] \delta \omega\,.
\end{align}
For the dilaton perturbation $f$\,, we have
\begin{align}
	\delta f = \Bigl\{ & \!\! - \tfrac{1}{4} \gamma \ls (\gamma+1) k^2 s_{A'}{}^{A'} - 2 k^{A'} k^{B'} s_{A'B'} \rs \notag \\[2pt]
		& + \gamma \, k^{A'} ( p \, t_{A'} + \overline{p} \, \overline{t}_{A'} ) - \tfrac{1}{2} \gamma (\gamma+1) \, k^2 t' + (\gamma-1) \, k^2 f \Bigr\} \, \delta \omega\,.
\end{align}
\end{subequations}

It is useful to rewrite the above results of beta-function \eqref{eq:lightconebeta} in terms of $s_{\mu\nu}$ and $a_{\mu\nu}$\,, together with $t_\mu{}^A$ defined as follows:
\begin{align}
	t_\mu{}^0 & \equiv \frac{1}{2} \bigl( t_\mu + \overline{t}_\mu \bigr)\,,
		&%
	t_\mu{}^1 & \equiv \frac{1}{2} \bigl( t_\mu - \overline{t}_\mu \bigr)\,.
\end{align}
We also write $p$ and $\overline{p}$ in terms of $p_0$ and $p_1$ as in \eqref{eq:pbarp}. Then, \eqref{eq:lightconebeta} becomes
\begin{subequations} \label{eq:Weylft}
\begin{align}
	\delta t' & = \bigl( - k^2 t_A{}^A + k^\Ap p_A t_\Ap{}^A \bigr) \, \delta \omega\,, \\[5pt]
	\delta{t} + \delta\overline{t} & = k^\Ap \bigl( p_0 \, t_\Ap{}^0 - k_\Ap t_0{}^0 - p_1 t_\Ap{}^1 + k_\Ap t_1{}^1 \bigr) \, \delta \omega \,,  \\[5pt]
	\delta{t} - \delta\overline{t} & = k^\Ap \bigl( p_0 \, t_\Ap{}^1 - k_\Ap t_0{}^1 - p_1 t_\Ap{}^0 + k_\Ap t_1{}^0 \bigr) \, \delta \omega \,,  \\[5pt]
	\delta{t}_{\Ap}{}^{A} & = \bigl( - k^2 t_{\Ap}{}^{A} + k_\Ap \, k^\Bp t_{\Bp}{}^{A} \bigr) \, \delta \omega\,, 
\end{align}
and
\begin{align}
	\delta s^{}_{\Ap \Bp} & =  
		\ls - 2 \ell^R_{\Ap\Bp} + 4 \, k^{}_\Ap k^{}_\Bp \, \ell^\Phi - (\gamma+1) k^{}_\Ap k^{}_\Bp \bigl( 2 \, t^A{}_A + s^\Cp{}_{\!\Cp} \bigr) \rs \delta \omega\,, \label{eq:R(ij)}\\[5pt]
	\delta a^{}_{\Ap \Bp} & = \ls i k^\Cp \ell^\CH_{\Ap\Bp\Cp} - \epsilon^{}_{AB} \, p^A \bigl( k^{}_\Ap t^{}_\Bp{}^B - k^{}_\Bp t^{}_\Ap{}^B \bigr) \rs \delta \omega \,, \\[5pt]
	\delta {\theta}^{}_{\!A\Ap } & = \Bigl[ - \ell^R_{A\Ap} + 2 \, p^{}_A k^{}_\Ap \ell^\Phi + \tfrac{i}{2} k^\Bp \epsilon^{}_A{}^B \ell^\CH_{B\Ap\Bp} \notag \\[2pt]
		& \quad - \tfrac{1}{2} (\gamma+1) \, p^{}_A k^{}_\Ap ( 2 \, t^B{}_B + s^\Bp{}^{}_\Bp ) + \tfrac{1}{2} \, p^{}_A \bigl( p^{}_B t^{}_\Ap{}^B - k^{}_\Ap t^{}_B{}^B \bigr) \Bigr] \delta \omega\,,  \\[5pt]
	\delta {\theta} & = \Bigl[ \tfrac{1}{2} \eta^{AB} \ell^R_{AB} - p^A p_A \ell^\Phi + \tfrac{i}{4} k^\Ap \epsilon^{AB} \ell^\CH_{\Ap AB} \notag \\
		& \hspace {4cm} + \tfrac{1}{4} (\gamma+1) \, p^B p_B \bigl( 2 t^A{}_A + s^\Ap{}_{\!\Ap} \bigr) \Bigr] \delta \omega\,, \\[5pt]
	\delta{f} & = \Bigl[ - \tfrac{1}{2} \gamma \, \delta^{\Ap\Bp}\ell^R_{\Ap\Bp} + (\gamma - 1) k^2 \ell^\Phi \notag \\[2pt]
		& \hspace{6.2mm} -  \tfrac{1}{4} (\gamma^2 - 1) k^2 \bigl( 2 t^A{}_A + s^\Ap{}_{\!\Ap} \bigr) + \tfrac{1}{2} \gamma \bigl( k^\Ap p_A t_\Ap{}^A - k^2 \, t_A{}^A \bigr) \Bigr] \delta \omega\,, \label{eq:deltaf}
\end{align}
\end{subequations}
where
\begin{subequations}
\begin{align}
	\ell^R_{\mu\nu} & \equiv - \tfrac{1}{2} \bigl( k^\Ap K^{}_\mu s^{}_{\!\Ap \nu} + k^\Ap K^{}_\nu s^{}_{\!\Ap \mu} - k^2 s^{}_{\mu\nu} - K^{}_\mu K^{}_\nu s^\Ap{}^{}_{\!\Ap} \bigr) \notag \\[2pt]
		& \quad - \tfrac{1}{2} \Bigl[ \delta_\mu^B \bigl( K_\nu t_{AB} - p_A t_{\nu B} \bigr) + \delta_\nu^B \bigl( K_\mu t_{AB} - p_A t_{\mu B} \bigr) \Bigr] p^A \notag \\[2pt]
		& \quad - \tfrac{1}{2} \Bigl[ K_\mu \bigl( p_A t_{\nu}{}^A - K_\nu t_A{}^A \bigr) + K_\nu \bigl( p_A t_\mu{}^A - K_\mu t_A{}^A \bigr) \Bigr]\,, \\[4pt]
	\ell^\CH_{\mu\nu\rho} & \equiv i \bigl( K_\mu a_{\nu\rho} + K_\rho a_{\mu\nu} + K_\nu a_{\rho\mu} \bigr)\,, \\[4pt]
	\ell^\Phi & \equiv f + \tfrac{1}{2} t_A{}^A + \tfrac{1}{4} s_\Ap{}^\Ap\,.
\end{align}
\end{subequations}
Here, we introduced the collective notation $K_\mu \equiv (p_A, k_\Ap)$\,.
  Weyl invariance enforces that all expressions in \eqref{eq:Weylft} vanish. 

\section{Nonrelativistic String Theory in Curved Spacetime} \label{sec:nstcurved}

In this section, we relate the Weyl transformations of the first excited vertex operators in \eqref{eq:Weylft} to the one-loop beta-functions in nonrelativistic string theory, linearized in perturbations around flat spacetime. We will later covariantize this linear result to include one-loop contributions to the beta-functions that are nonlinear in perturbations around flat spacetime. Setting these beta-functions to zero at the leading $\alpha'$-order, we find the equations of motion that govern the dynamics of the string Newton-Cartan background, $B$-field and dilaton.

\subsection{Strings in a String Newton-Cartan Background} \label{sec:ingrecurve}

First, we review the basic ingredients of the nonrelativistic string theory sigma model in a string Newton-Cartan background coupled to the Kalb-Ramond and dilaton field.
The free nonrelativistic string theory action in flat spacetime $S_\text{f}$ in \eqref{eq:actionflatcg} is invariant under the following global transformations:
\be
	\delta x^A = \Xi^A - \Lambda \, \epsilon^A{}_B x^B\,,
		\qquad
	\delta x^{\Ap} = \Xi^{\Ap} - \Lambda^{\Ap}{}_{\Bp } \, x^{\Bp } + \Lambda_A{}^{\Ap} x^A\,,
\ee
accompanied by appropriate transformations of $\lambda$ and $\overline{\lambda}$\,. The associated generators in the global symmetry algebra are as follows:
\begin{align*}
	\Xi^A: & \quad \text{longitudinal translations} && \hspace{-2.5cm} H_A \\
	\Xi^{\Ap}: & \quad \text{transverse translations} && \hspace{-2.5cm} P_{\Ap} \\
	\Lambda: & \quad \text{longitudinal Lorentz rotation} && \hspace{-2.5cm} M \\
	\Lambda^{\Ap \Bp }: & \quad \text{transverse spatial rotations} && \hspace{-2.5cm} J_{\Ap \Bp } \\
	\Lambda^{A\Ap }: & \quad \text{string Galilean boosts} && \hspace{-2.5cm} G_{A\Ap }
\end{align*}
Moreover, the Poisson bracket between $G_{A\Ap }$ and $P_{\Ap }$ gives a noncentral extension $Z_A$\,: 
\be
	[G_{A\Ap }, P_{\Bp }] = \delta_{\Ap \Bp } Z_A\,.
\ee 
This extension exists because the Lagrangian associated with the action \eqref{eq:actionflatcg} is invariant under the string Galilean boosts up to a boundary term \cite{nrGalilei, stringyNC}. There are other noncentral extensions in the full global symmetry algebra \cite{stringyNC, Harmark:2018cdl, stringyNClimit}, which we are not interested for the purpose of this paper. The global symmetry algebra that consists of the aforementioned generators, including all noncentral extensions, is dubbed the string Newton-Cartan algebra.\,\footnote{We follow the terminology introduced in \cite{Bergshoeff:2018vfn, stringyNClimit}. In older literature, for examples in \cite{stringyNC, Bergshoeff:2018yvt}, this algebra is referred to as the ``extended string Galilei algebra."}

The appropriate spacetime geometry that the nonrelativistic strings coupled to is the string Newton-Cartan geometry, which realizes the string Newton-Cartan algebra as a gauge symmetry acting on the target space \cite{nrGalilei, stringyNC, Bergshoeff:2018yvt, stringyNClimit}.\,\footnote{More precisely, the full spacetime symmetry algebra realized in nonrelativistic string theory is an extension of the string Newton-Cartan algebra originally introduced in \cite{nrGalilei}. See \cite{stringyNClimit, Harmark:2018cdl} for further discussions.} The string Newton-Cartan geometry is defined as follows. Let $\mathcal{T}_p$ be the tangent space attached to a point $p$ in the spacetime $\mathcal{M}$\,. We decompose $\mathcal{T}_p$ into two longitudinal directions indexed by $A = 0, 1$ and $d-2$ transverse directions indexed by $\Ap = 2, \cdots, d-1$\,, respectively. We introduce the longitudinal Vielbein fields $\tau_\mu{}^A$ and the transverse Vielbein fields $E_\mu{}^\Ap$\,, which in the flat limit become $\tau_\mu{}^A \rightarrow \delta_\mu^A$ and $E_\mu{}^\Ap \rightarrow \delta_\mu^\Ap$\,.
The inverse Vielbein fields $\tau^\mu{}_A$ and $E^\mu{}_{\Ap }$ are defined via the following relations:
\begin{subequations}
\begin{align}
	\tau^\mu{}_A \tau_\mu{}^B & = \delta^B_A\,,
		&
	\hspace{-2cm}\tau_\mu{}^A \tau^\nu{}_A + E_\mu{}^{\Ap} E^\nu{}_{\Ap} & = \delta_\mu^\nu\,, \\[2pt]
	E_\mu{}^{\Ap} E^\mu{}_{\Bp} & = \delta^{\Ap}_{\Bp}\,,
		&
	\hspace{-2cm}\tau^\mu{}_A E_\mu{}^{\Ap} = E^\mu{}_{\Ap} \tau_\mu{}^A & = 0\,. 
\end{align}
\end{subequations}
Moreover, associated with the noncentral extension $Z_A$ we introduce a gauge field $m_\mu{}^A$\,, whose flat limit is $m_\mu{}^A \rightarrow 0$\,. 

In the presence of a Kalb-Ramond field $B_{\mu\nu}$ and a dilaton field $\Phi$\,, the sigma model of nonrelativistic string theory on an arbitrary string Newton-Cartan geometry is given by \cite{Bergshoeff:2018yvt}
\begin{align} \label{eq:originalcurved}
	S & = \frac{1}{4\pi\alpha'} \int d^2 \sigma \sqrt{h} \bigl( h^{\alpha\beta} \p_\alpha x^\mu \p_\beta x^\nu H_{\mu\nu} + \lambda \, \overline{\qd} x^\mu \tau_\mu + \overline{\lambda} \, \qd x^\mu \overline{\tau}_\mu \bigr) \notag \\
	& \quad + \frac{1}{4\pi\alpha'} \int d^2 \sigma \, \epsilon^{\alpha\beta} \p_\alpha x^\mu \p_\beta x^\nu B_{\mu\nu} + \frac{1}{4\pi} \int d^2 \sigma \sqrt{-h} \, R^{(2)} \, \Phi\,, 
\end{align}
where $R^{(2)}$ is the scalar curvature of $h_{\alpha\beta}$ and 
\begin{subequations} \label{eq:tautauH}
\begin{align}
	\tau_\mu & = \tau_\mu{}^0 + \tau_\mu{}^1\,,
		\qquad
	\overline{\tau}_\mu = \tau_\mu{}^0 - \tau_\mu{}^1\,, \\[2pt]
	H_{\mu\nu} & = E_\mu{}^\Ap E_\nu{}^\Bp \delta_{\Ap \Bp} + \bigl( \tau_\mu{}^A m_\nu{}^B + \tau_\nu{}^A m_\mu{}^B \bigr) \eta_{AB}\,.
\end{align}
\end{subequations}
Note that $\tau_\mu{}^A$ and $H_{\mu\nu}$ are all the quantities with covariant curved spacetime indices that are invariant under the string Galilean boosts.
Further note that $m_\mu{}^A$ transforms nontrivially under the $Z_A$ transformation, which we parametrize by $\sigma^A$:\,\footnote{This $\sigma^A$ shall not be confused with the worldsheet coordinates $\sigma^\alpha$\,.} 
\be
	\delta m_\mu{}^A = D_\mu \sigma^A\,,
\ee 
where the derivative $D_\mu$ is covariant with respect to the longitudinal Lorentz boost transformations acting on the index $A$\,. In order for this $\sigma^A$ transformation to be a gauge symmetry of the action in \eqref{eq:originalcurved}, first, one has to transform $\lambda$ and $\overline{\lambda}$ as
\be
	\delta \lambda = \mathcal{D} x^\mu D_\mu \overline{\sigma}\,,
		\qquad
	\delta \overline{\lambda} = \overline{\mathcal{D}} x^\mu D_\mu \sigma\,,
\ee
where $\sigma \equiv \sigma^0 + \sigma^1$ and $\overline{\sigma} \equiv \sigma^0 - \sigma^1$\,;
second, one also has to impose the hypersurface orthogonality condition in the two-dimensional foliation structure on $\mathcal{M}$ \cite{Bergshoeff:2018yvt, stringyNClimit},\,\footnote{This is an analogue of the hypersurface orthogonality condition in spacetime with a one-dimensional foliation sturcture.}
\be \label{eq:eq:DtauA00}
	D_{[\mu} \tau_{\nu]}{}^A = 0 
		\quad \implies \quad
	\p_{[\mu} \tau_{\nu]}{}^A = \Omega_{[\mu}{}^{AB} \, \tau_{\nu]B}\,,
\ee
where $\Omega_\mu{}^{AB}$ denotes the spin connection associated with the longitudinal Lorentz rotation symmetry. There are $d$ components in \eqref{eq:eq:DtauA00} that can be used to solve for $\Omega_\mu{}^{AB}$ in terms of $\tau_\mu{}^A$\,. 
The remaining $d(d-2)$ equations can be rewritten in a compact form,
\be \label{eq:DtauA0}
	\epsilon^{(A}{}^{}_C \, \tau_{[\mu}{}^{B)} \, \p^{}_{\nu\phantom{]}\!} \tau^{}_{\rho]}{}^C = 0\,.
\ee	
This $Z_A$ symmetry prohibits the $\lambda \overline{\lambda}$ operator from being generated in \eqref{eq:originalcurved}. Otherwise, the $\lambda \overline{\lambda}$ operator would have deformed nonrelativistic string theory towards relativistic string theory.

The path integral for the action \eqref{eq:originalcurved} is given by
\be \label{eq:pathintegral}
	\mathcal{Z} = \int \mathscr{D} \lambda \, \mathscr{D} \overline{\lambda} \, \mathscr{D} x^\mu \sqrt{G} \exp(-S) = \int \mathscr{D} \lambda \, \mathscr{D} \overline{\lambda} \, \mathscr{D} x^\mu \exp(-S_G)\,,
\ee
where
\be
	G \equiv \det^{(d)} \bigl( H_{\mu\nu} \bigr) \, \det^{(2)} \bigl( \tau_\rho{}^A H^{\rho\sigma} \tau_\sigma{}^B \bigr)\,,
\ee
and the effective action $S_G$ is defined to be
\begin{align} \label{eq:SG}
	S_G & \equiv \frac{1}{4\pi\alpha'} \int d^2 \sigma \sqrt{h} \bigl( \, h^{\alpha\beta} \p_\alpha x^\mu \p_\beta x^\nu H_{\mu\nu} + \lambda \, \overline{\qd} x^\mu \tau_\mu + \overline{\lambda} \, \qd x^\mu \overline{\tau}_\mu \, \bigr) \notag \\
	& \quad + \frac{1}{4\pi\alpha'} \int d^2 \sigma \, \epsilon^{\alpha\beta} \p_\alpha x^\mu \p_\beta x^\nu B_{\mu\nu} + \frac{1}{4\pi} \int d^2 \sigma \sqrt{-h} \, R^{(2)} \, \bigl( \Phi - \tfrac{1}{4} \ln G \, \bigr)\,.
\end{align}
Note that $\lambda$ and $\overline{\lambda}$ play the role of a Lagrange multiplier in \eqref{eq:SG}.   These Lagrange multipliers are not uniquely defined; their redefinitions generate symmetry transformations that relate different string Newton-Cartan geometries. Indeed, apart from the target space gauge symmetries in string Newton-Cartan gravity without any Kalb-Ramond or dilaton field, we observe additional St\"{u}ckelberg-type symmetries in \eqref{eq:SG} by also allowing the Kalb-Ramond and dilaton field to transform as follows:\,\footnote{When $C = \overline{C}$\,, the transformations in \eqref{eq:lltrans} are the ones that keep the Buscher rules associated with the longitudinal spatial T-duality invariant. See \cite{Bergshoeff:2018yvt} for more details. A special form of the transformations in \eqref{eq:lltrans} also found use in \cite{Harmark:2018cdl}.}
\begin{subequations} \label{eq:lltranslambda}
\begin{align}
	\lambda & \rightarrow {C}^{-1} \bigl( \lambda - \qd x^\mu \, \overline{C}_\mu \bigr)\,,
		&%
	\overline{\lambda} & \rightarrow {\overline{C}}^{-1} \bigl( \overline{\lambda} - \overline{\qd} x^\mu C_\mu \bigr)\,, 
\end{align}
together with
\begin{align} \label{eq:lltrans}
	\tau_\mu & \rightarrow C \, \tau_\mu\,, 
		&%
	H_{\mu\nu} & \rightarrow H_{\mu\nu} - \bigl( C_\mu{}^A \tau_\nu{}^B + C_\nu{}^A \tau_\mu{}^B \bigr) \, \eta_{AB}\,, \\[2pt]
	\overline{\tau}_\mu & \rightarrow \overline{C} \, \overline{\tau}_\mu\,, 
		&%
	B_{\mu\nu} & \rightarrow B_{\mu\nu} + \bigl( C_\mu{}^A \tau_\nu{}^B - C_\nu{}^A \tau_\mu{}^B \bigr) \, \epsilon_{AB}\,,
\end{align}
\end{subequations}
and
\begin{align} \label{eq:lltransPhi}
	\Phi \rightarrow \Phi + \frac{1}{2} \ln \bigl( {C} \, \overline{C} \bigr)\,.
\end{align}
We defined
\be
	C_\mu = C_\mu{}^0 + C_\mu{}^1\,,
		\qquad
	\overline{C}_\mu = C_\mu{}^0 - C_\mu{}^1\,.
\ee
Here, $C_\mu{}^A$ is an arbitrary function of $x^\mu$ but $C$ and $\overline{C}$ are constrained as follows in order to preserve the geometric constraints \eqref{eq:DtauA0}:
\be \label{eq:condCbarCfull}
	E^\mu{}_\Ap \p_\mu \bigl( C \overline{C} \bigr) = 0\,.
\ee
To linear order, we simply have
\be \label{eq:pApC}
	\p_\Ap \bigl( C \overline{C} \bigr) = 0\,,
\ee
i.e., $C \overline{C}$ is independent of $x^\Ap$\,.

\subsection{Linearized Beta-Functions from Weyl Transformations} \label{sec:lbWt}

Note that the action $S_G$ in \eqref{eq:SG} consists of the most general  operators that are classically invariant under the global Weyl transformations and carry zero winding. Moreover, $S_G$ is classically invariant under the local Weyl transformation except for the dilaton term, which can be cancelled by including the quantum contributions to the Weyl transformations. In the limit that $H_{\mu\nu}$ is close to $\delta_\mu^\Ap \delta_\nu^\Bp \delta_{\Ap\Bp}$\,, $\tau_\mu{}^A$ is close to $\delta_\mu^A$\,, and both $B_{\mu\nu}$ and $\Phi$ are small, we take the following expansions:
\begin{subequations} \label{eq:Fourier}
\begin{align}
	\tau_\mu{}^A (x) & = \delta_\mu^A + T_\mu{}^A (x)\,, 
		&%
	\hspace{-1cm} T_\mu{}^A (x) & \equiv - 4 \pi g \int d^d K \, t_{\mu}{}^A (K) \, e^{i K_\mu x^\mu}\,, \label{eq:Fouriertau} \\
	H_{\mu\nu} (x) & = \delta_\mu^\Ap \delta_\nu^\Bp \delta_{\Ap\Bp} + S_{\mu\nu} (x)\,, 
		&%
	\hspace{-1cm} S_{\mu\nu} (x) & \equiv - 4 \pi g \int d^d K \, s_{\mu\nu} (K) \, e^{i K_\mu x^\mu}\,, \\
	B_{\mu\nu} (x) & = - 4 \pi g \int d^d K \, a_{\mu\nu} (K) \, e^{i K_\mu x^\mu}\,, \\
	\Phi (x) & = - 4 \pi g \, \int d^d K \, \phi (K) \, e^{i K_\mu x^\mu}\,,
\end{align}
\end{subequations}
where $K_\mu \equiv (p^A, k^{\Ap })$\,.
This leads to the expansion of the path integral in \eqref{eq:pathintegral},
\be \label{eq:SGexpand}
	\mathcal{Z} = \int \mathscr{D} \lambda \, \mathscr{D} \overline{\lambda} \, \mathscr{D} x^\mu \, e^{-S_f} + \int d^d K \, \mathcal{V} (K) + \cdots\,,
\ee
where $S_f$ is the flat spacetime action given in \eqref{eq:actionflat} and $\mathcal{V} (K)$ is given by \eqref{eq:fevobefore} with $u = 0$\,. The Weyl transformation of $\mathcal{Z}$ gives the Weyl anomaly
\be
	\delta_W \mathcal{Z} = - \frac{1}{2\pi} \int d^2 \sigma \sqrt{h} \, \delta \omega \, \langle T^\alpha{}_\alpha \rangle\,, 
\ee
where the trace of the stress-energy tensor $T_{\alpha\beta}$ is given by
\begin{align}
	T^\alpha{}_{\alpha} & = - \frac{1}{2\alpha'} \Bigl[ \, ( \beta^H_{\mu\nu} + \beta^B_{\mu\nu} ) \, \qd X^\mu \overline{\qd} X^\nu + \beta^\tau_\mu \, \lambda \, \overline{\mathcal{D}} x^\mu + \beta^{\overline{\tau}}_\mu \, \overline{\lambda} \, {\mathcal{D}} x^\mu + \alpha' \beta^F R \, \Bigr]\,.
\end{align}
This defines the beta-functions for the background fields.
Here, for example, $\beta^H_{\mu\nu}$ denotes the beta-function of $H_{\mu\nu}$\,. Define
\begin{subequations} \label{eq:defineThetaF}
\begin{align}
	\Theta & \equiv - \tfrac{1}{4} \tau^\mu{}_A \tau^\nu{}_B \bigl( \eta^{AB} H_{\mu\nu} - \epsilon^{AB} B_{\mu\nu} \bigr)\,, \\[8pt]
	\Theta_{A\Ap} & \equiv \tfrac{1}{2} \tau^\mu{}_B E^\nu{}_\Ap \bigl( \delta^B_A H_{\mu\nu} + \epsilon_{A}{}^B B_{\mu\nu} \bigr)\,, \\[2pt]
	F & \equiv \Phi - \tfrac{1}{4} \ln \Bigl[ \det^{(d)}\bigl(H_{\mu\nu}\bigr) \det^{(2)} \bigl( \tau_\rho{}^A H^{\rho\sigma} \tau_\sigma{}^B\bigr) \Bigr]\,.
\end{align}
\end{subequations}
Note that
\be
	F (x) = - 4 \pi g \, \int d^d K \, f (K) \, e^{i K_\mu x^\mu}\,,
\ee
where $f$ matches the one introduced in \eqref{eq:fevo}. We further define
\begin{subequations}
\begin{align}
	\beta^\tau_{AB} & \equiv \tfrac{1}{2} \bigl( \tau^\mu{}^{}_A \, \beta^\tau_{\mu B} + \tau^\mu{}^{}_B \, \beta^\tau_{\mu A} \bigr)\,,
		&%
	\beta^\Theta_{A\Ap} & \equiv \tfrac{1}{2} \tau^\mu{}_B E^{\nu}{}^{}_\Ap \bigl( \delta_A^B \, \beta^H_{\mu\nu} + \epsilon_A{}^B \, \beta^B_{\mu\nu} \bigr)\,, \\[4pt]
	\beta^\tau_{\Ap A} & \equiv E^\mu{}_{\Ap} \, \beta^\tau_{\mu A}\,, 
		&%
	\beta^\Theta & \equiv - \tfrac{1}{4} \tau^\mu_A \tau^\nu{}_B \bigl( \eta^{AB} \beta^H_{\mu\nu} - \epsilon^{AB} \beta^B_{\mu\nu} \bigr)\,, \\[4pt]
	\beta^H_{\Ap\Bp} & \equiv E^\mu{}_\Ap E^\nu{}_\Bp \beta^H_{\mu\nu}\,, 
		&%
	\beta^B_{\Ap\Bp} & \equiv E^\mu{}_\Ap E^\nu{}_\Bp \beta^B_{\mu\nu}\,,
\end{align}
\end{subequations}
where $\beta^\tau_{\mu A}$ is given by
\be
	\beta_{\mu0}^\tau \equiv - \tfrac{1}{2} \bigl( \beta^\tau_\mu + \beta^{\overline{\tau}}_\mu \bigr)\,,
		\qquad
	\beta_{\mu1}^\tau \equiv \tfrac{1}{2} \bigl( \beta^\tau_\mu - \beta^{\overline{\tau}}_\mu \bigr)\,.
\ee
It is also convenient to adopt a renormalization scheme with $\gamma = -1$\,. Using the above definitions, from \eqref{eq:Weyltransformations} we read off the beta-functions to linear order in $T_\mu{}^A (x)$\,, $S_{\mu\nu} (x)$\,, $B_{\mu\nu} (x)$ and $\Phi (x)$\,,
\begin{subequations} \label{eq:deltatmuA}
\begin{align}
	\beta^\tau_{AB} & \approx - \frac{\alpha'}{4} \p^\Ap \bigl( \p_{\Ap} \tau_{AB} - \p_{A} \tau_{\Ap B} \bigr) + (A \leftrightarrow B) \,, \\[2pt]
	\beta^\tau_{\Ap A}
		& \approx - \frac{\alpha'}{2} \p^\Bp \bigl( \p_{\Bp} \tau_{\Ap A} - \p_{\Ap} \tau_{\Bp A} \bigr)\,,
\end{align}
\end{subequations}
and 
\begin{subequations} \label{eq:betafunction}
\begin{align}
	\beta^H_{\Ap \Bp} & \approx \alpha' \bigl( L^R_{\Ap\Bp} + 2 \, \p^{}_\Ap \p^{}_\Bp \Phi \bigr)\,, \\[5pt]
	\beta^B_{\Ap \Bp} & \approx - \frac{\alpha'}{2} \Bigl[ \p^\Cp \CH^{}_{\Cp\Ap\Bp} + \epsilon^{}_{AB} \bigl( \p^A \p^{}_{\Ap} \tau^{}_{\Bp}{}^B - \p^A \p^{}_{\Bp} \tau^{}_{\Ap}{}^B \bigr) \Bigr] \,, \\[5pt]
	\beta^\Theta_{A\Ap } & \approx \frac{\alpha'}{2} \, \Bigl[ L^R_{A\Ap} + 2 \, \p^{}_A \p^{}_\Ap \Phi - \tfrac{1}{2} \, \epsilon_A{}^B \p^\Bp \CH^{}_{\Ap \Bp B} + \tfrac{1}{2} \, \p^{}_A \bigl( \p^{}_{B} \tau^{}_{\Ap}{}^B - \p^{}_{\Ap} \tau^{}_{B}{}^B \bigr) \Bigr] \,, \label{eq:RAi0}\\[5pt]
	\beta^\Theta & \approx - \frac{\alpha'}{4} \bigl( \eta^{AB} L^R_{AB} + 2 \p^A \p^{}_A \Phi  + \tfrac{1}{2} \, \epsilon^{AB} \p^\Ap \CH^{}_{\Ap AB} \bigr)\,,\label{eq:R0} \\[5pt]
	\beta^F & \approx \frac{c}{6} - \alpha' \Bigl[ \p^{}_\Ap \p^\Ap \Phi + \tfrac{1}{4} \, \delta^{\Ap\Bp} L^R_{\Ap\Bp} 
		+ \tfrac{1}{4} \, \p^\Ap \bigl( \p^{}_{A} \tau^{}_{\Ap}{}^A - \p^{}_{\Ap} \tau^{}_{A}{}^A \bigr) \Bigr]\,, \label{eq:betaF}
\end{align}
\end{subequations}
where
\begin{subequations}
\begin{align}
	L^R_{\mu\nu} & \equiv \tfrac{1}{2} \bigl( \p^\Ap \p_\mu H_{\Ap \nu} + \p^\Ap \p_\nu H_{\Ap \mu} - \p^\Ap \p_\Ap H_{\mu\nu} - \p_\mu \p_\nu H^\Ap{}_\Ap \bigr) \notag \\[3pt]
		& \quad + \tfrac{1}{2} \Bigl[ \delta_\mu^B \p^A \bigl( \p_\nu \tau_{AB} - \p_A \tau_{\nu B} \bigr) + \delta_\nu^B \p^A \bigl( \p_\mu \tau_{AB} - \p_A \tau_{\mu B} \bigr) \notag \\
		& \quad\qquad\quad\,\,\,\,\, + \p_\mu \bigl( \p_A \tau_\nu{}^A - \p_\nu \tau_A{}^A \bigr) + \p_\nu \bigl( \p_A \tau_\mu{}^A - \p_\mu \tau_A{}^A \bigr) \Bigr ]\,, \label{eq:LR} \\[5pt]
	\CH_{\mu\nu\rho} & \equiv \p_\mu B_{\nu\rho} + \p_\rho B_{\mu\nu} + \p_\nu B_{\rho\mu}\,.
\end{align}
\end{subequations}
When a spacetime index is contracted with the spacetime index of an (inverse) Vielbein field with no derivatives acting on it, we replace this spacetime index with the flat index of this (inverse) Vielbein field. For example,
\be
	\p_B H_{A\Ap } = \tau^\mu{}_A \, \tau^\rho{}_B \, E^\nu{}_\Ap \, \p_\rho H_{\mu\nu}\,.
\ee
Since we are focusing on the linear order terms, a contraction of an index $\mu$ with $\tau^\mu{}_A$ or $E^\mu{}_\Ap$ is usually reduced to a contraction with $\delta^\mu_A$ or $\delta^\mu_\Ap$\,.

The central charge $c$ in \eqref{eq:betaF} comes from the flat spacetime anomaly, which contains a contribution $c_\text{trans.} = d-2$ from the $d-2$ transverse fields $x^\Ap$\,, a contribution $c_\text{long.} = 2$ from the longitudinal commuting $\beta\gamma$ system and $c_\text{ghost} = - 26$ from the $bc$ ghost system just as in relativistic string theory. In total, 
\be \label{eq:centralcharge}
	c = c_\text{trans.} + c_\text{long.} + c_\text{ghost} = d - 26\,.
\ee
This is the same information that we already learnt from the BRST quantization in \S\ref{sec:renorm}, which simply declares that the critical dimension is 26.

It is important to note that, due to the path integral identities \eqref{eq:dXdxe}, \eqref{eq:bdXdxe} and \eqref{eq:addid}, not all background field perturbations introduced in \eqref{eq:Fourier} are nontrivial, as they  are   multiplied by a vanishing on-shell renormalized vertex operator. The only nontrivial background field perturbations are given in \eqref{eq:fevo}. Therefore, it is only physically meaningful to define beta-functions for these background fields. As a result, the only beta-functions that we can define are for $\tau_{AB}$\,, $\tau_{A'A}$\,, $H_{A'B'}$\,, $B_{A'B'}$\,, $\Theta$\,, $\Theta_{AA'}$ and $F$ as in \eqref{eq:deltatmuA} and \eqref{eq:betafunction}, which are invariant under the transformations given in \eqref{eq:lltranslambda}.

\subsection{Geometric Constraints and Weyl Invariance} \label{sec:geocons}

In this subsection, we study the Weyl invariance requirement from setting the beta-functions of (components of) the longitudinal Vielbein field $\tau_\mu{}^A$ in \eqref{eq:deltatmuA} to zero, which results in the following geometric constraints:
\begin{align} \label{eq:betataumuA2}
	\p^\Ap \bigl( \p_{\Ap} \tau_{AB} - \p_{A} \tau_{\Ap B} \bigr) + (A \leftrightarrow B) = 0\,, 
		\qquad
	\p^\Bp \bigl( \p_{\Bp} \tau_{\Ap A} - \p_{\Ap} \tau_{\Bp A} \bigr) = 0\,.
\end{align}
We will show that the constraints in \eqref{eq:betataumuA2} are equivalent to the linearization of the hypersurface orthogonality condition in \eqref{eq:DtauA0}.

First, using \eqref{eq:Fouriertau}, we take the Fourier transform of the equations in \eqref{eq:betataumuA2}, which gives
\begin{align} \label{eq:kApgc}
	k^\Ap \bigl( k_{\Ap} t_{AB} - p_A t_{\Ap B} \bigr) + (A \leftrightarrow B) = 0\,,
		\qquad%
	k^\Bp \bigl( k_{\Bp} t_{\Ap}{}^A - k_{\Ap} t_{\Bp}{}^A \bigr) = 0\,,
\end{align}
i.e. the equations that follow from setting $\delta t' = \delta t = \delta \overline{t} = \delta t_{\Ap}{}^A = 0$ in \eqref{eq:Weylft}. Note that these equations hold for all $k_\Ap$ and, in particular, $k_\Ap \neq 0$ for off-shell intermediate zero winding string states that mediate (string) Newtonian gravitational force among winding strings. Due to the rotational invariance in the transverse sector, without losing generality, we are free to choose a coordinate system in which
\be \label{eq:specialkAp}
	k_\Ap = ( \, k_2, \underbrace{0, \cdots, 0}_{d-3} \, )\,,
		\qquad
	k \neq 0\,.
\ee
In these coordinates, \eqref{eq:kApgc} becomes
\be \label{eq:k2t}
	k_2 \bigl( t_{AB} + t_{BA} \bigr) = p_A t_{2B} + p_B t_{2A}\,, 
		\qquad
	k_I \, t_2{}^A = k_2 \, t_I{}^A\,,
		\qquad
	I = 3, 4, \cdots, d-1\,. 
\ee
The inverse Fourier transform of \eqref{eq:k2t} gives
\be \label{eq:gcstrong}
	\bigl( \p_{\Ap} \tau_{AB} - \p_{A} \tau_{\Ap B} \bigr) + (A \leftrightarrow B) = 0\,, 
		\qquad
	\p_{\Bp} \tau_{\Ap A} - \p_{\Ap} \tau_{\Bp A} = 0\,.
\ee
Another way to put this is that, in order to preserve the rotational symmetry in the transverse sector in \eqref{eq:betataumuA2}, the expressions on the left hand side of the equations in \eqref{eq:gcstrong} cannot be constant vectors with respect to the index $\Ap$ and hence have to vanish. Therefore, by requiring rotational invariance in the transverse sector, we see that \eqref{eq:betataumuA2} leads to \eqref{eq:gcstrong}.\,\footnote{This argument does not apply to \eqref{eq:pApC}: since a scalar $C \overline{C}$ automatically preserves the transverse rotational symmetry, at the linear order, \eqref{eq:pApC} only constrains $C \overline{C}$ to be a constant in $x^\Ap$ instead of identically zero.} 
Finally, we note that the equations in \eqref{eq:gcstrong} are linearized expressions for  
\be \label{eq:DtauA02}
	\epsilon^{(A}{}^{}_C \, \tau_{[\mu}{}^{B)} \, \p^{}_{\nu\phantom{]}\!} \tau^{}_{\rho]}{}^C = 0\,, 
\ee
i.e. the constraint equation \eqref{eq:DtauA0}. This implies that the hypersurface orthogonality condition \eqref{eq:eq:DtauA00} leads to the vanishing beta-function for $\tau_\mu{}^A$ at linear order (and higher, see below).

\subsection{Extension to Higher Order Terms} \label{sec:nonlinear}

In principle, the method used to calculate the linear order terms in the equations of motion can also be applied to higher order perturbations in $T_\mu{}^A$\,, $S_{\mu\nu}$\,, $B_{\mu\nu}$ and $\Phi$ introduced in \eqref{eq:Fourier}. We will still focus on the contributions to the beta-functions that are linear in $\alpha'$\,. To calculate the second order terms in $T_\mu{}^A$\,, $S_{\mu\nu}$\,, $B_{\mu\nu}$ and $\Phi$\,, one needs to keep the following quadratic term in the path integral \eqref{eq:SGexpand},
\be \label{eq:Z(2)}
	\mathcal{Z}^{(2)} = \frac{1}{2} \int d^d K \, d^d K' \, \mathcal{V} (K) \, \mathcal{V} (K')\,.
\ee 
The Weyl transformation of $\mathcal{Z}^{(2)}$ contributes to nonlinear terms in beta-functions. This calculation is straightforward by evaluating various OPEs. However, since we already understand the underlying gauge symmetries of the target spacetime, it is more efficient to determine higher order terms in the equations of motion by requiring that all ingredients are covariant.

First, we collect all combinations of string Newton-Cartan fields that are invariant under the string Galilei boosts, namely, $E^\rho{}_\Ap$\,, $\tau_{\mu}{}^A$\,, $H_{\mu\nu}$ and $N^{\mu\nu}$\,,
where
\be
	N^{\mu\nu} \equiv \tau^{\mu}{}_A \tau^{\nu A} - \bigl( E^{\mu}{}_{\Ap} \tau^{\nu}{}_A + E^{\nu}{}_{\Ap} \tau^{\mu}{}_A \bigr) m_\lambda{}^A E^\lambda{}_{\Ap}\,.
\ee
In string Newton-Cartan geometry, there is a unique Christoffel symbol\,\footnote{To write the Christoffel symbol in the form of \eqref{eq:Christoffel}, one has to make a special choice of the variable $W_{ABA'}$ in \cite{stringyNClimit}. These extra variables in $W_{ABA'}$ drops out in the final equations of motion \cite{stringyNClimit}.}
\begin{align} \label{eq:Christoffel}
	\Gamma^\rho{}_{\mu\nu} & = \tfrac{1}{2} E^{\rho\sigma} \bigl( \p_\mu H_{\sigma\nu} + \p_\nu H_{\sigma\mu} - \p_\sigma H_{\mu\nu} \bigr) + \tfrac{1}{2} N^{\rho\sigma} \bigl( \p_\mu \tau_{\sigma\nu} + \p_\nu \tau_{\sigma\mu} - \p_\sigma \tau_{\mu\nu} \bigr)\,,
\end{align}
where we defined
\be
	E^{\mu\nu} \equiv E^\mu{}_{A'} E^\nu{}_{B'} \delta^{A'B'}\,,
		\qquad%
	\tau_{\mu\nu} \equiv \tau_\mu{}^A \tau_\nu{}^B \eta_{AB}\,.
\ee
With respect to the Christoffel symbol \eqref{eq:Christoffel}, we define the covariant derivative $\nabla_\mu$ and the Riemann tensor,
\be
	R^\rho{}_{\sigma\mu\nu} = \p_\mu \Gamma^\rho{}_{\nu\sigma} - \p_\nu \Gamma^\rho{}_{\mu\sigma} + \Gamma^\rho{}_{\mu\lambda} \Gamma^\lambda{}_{\nu\sigma} - \Gamma^\rho{}_{\nu\lambda} \Gamma^\lambda{}_{\mu\sigma}\,.
\ee
We also define the Ricci tensor $R_{\mu\nu} \equiv R^\rho{}_{\mu\rho\nu}$\,. 
One can show that $L^R_{\mu\nu}$ defined in \eqref{eq:LR} is nothing but the linearized $R_{\mu\nu}$\,.
These constitute all the covariant ingredients once we also include $D_{[\mu} \tau_{\nu]}{}^A$ that appears in the hypersurface orthogonality condition in \eqref{eq:eq:DtauA00}. Note that $\nabla_\mu \tau_{\nu\rho} = \nabla_\mu E^{\nu\rho} = 0$\,. 

Next, we consider the equations of motion from the vanishing beta-functions of $\tau_\mu{}^A$\,. At linear order, these are the geometric constraints \eqref{eq:gcstrong}, whose covariantization takes the form
\be
	D_{[\mu} \tau_{\nu]}{}^A = 0\,,
\ee
i.e. the hypersurface orthogonality condition in \eqref{eq:eq:DtauA00}.  
One might wonder whether higher order contributions  
to the beta-functions of $\tau_\mu{}^A$ sabotage the hypersurface orthogonality condition.
If true, this would make the $Z_A$ symmetry  anomalous. To address this question, instead of working directly with \eqref{eq:Z(2)}, we take the conformal gauge and the following expansions  around $x^\mu = 0$ (or, equivalently, around any other point $x_0^\mu$):
\begin{align}
	g_i (x) & = g_i (0) + x^\mu \lr \, \p_\mu g_i (x) \big|_{x^\nu = 0} \,\rr + \cdots\,, 
\end{align}
where $g_i$ denotes any of the couplings $\{ \tau_\mu{}^A\,, H_{\mu\nu}\,, B_{\mu\nu}\,, F \}$\,.
It is sufficient to focus on the contributions to the Weyl transformations $\delta_W g_i$ that take the form of $\p_\mu g_i \, \p_\nu g_j$\,, which requires analyzing the terms cubic in $\{ \, x^\mu, \lambda\,, \overline{\lambda} \, \}$ in the action. We can also set all higher derivative terms (containing two or more derivatives) of $g_i$ to zero in the calculation. Then, the full nonlinear expressions can be constructed by covariantization. 

We first collect all terms cubic in $\{ \, x^\mu, \lambda\,, \overline{\lambda} \, \}$ in the action,
\begin{align} \label{eq:S(3)0}
	S^{(3)} & = \frac{1}{4\pi\alpha'} \int dz \, d\overline{z} \, \Bigl[ \p_z x^\mu \p_{\overline{z}} x^\nu x^\rho \p_\rho (H_{\mu\nu} + B_{\mu\nu}) \notag \\
		& \hspace{2.5cm} + \lambda \, \p_{\overline{z}} x^\mu \bigl( x^\nu \p_\nu \tau_\mu \bigr) + \overline{\lambda} \, \p_{{z}} x^\mu \bigl( x^\nu \p_\nu \overline{\tau}_\mu \bigr)
		- 4 \alpha' \p_z \p_{\overline{z}} \omega \bigl( x^\mu \p_\mu F \bigr) \Bigr]\,.
\end{align}
Using integration by parts, we rewrite $S^{(3)}$ in \eqref{eq:S(3)0} as
\begin{align} \label{eq:S(3)}
	S^{(3)} & = \frac{1}{4\pi\alpha'} \int dz \, d\overline{z} \, \Bigl[ \p_z x^\mu \p_{\overline{z}} x^\nu x^\rho \bigl( \p_\rho H_{\mu\nu} + \tfrac{1}{3} \CH_{\mu\nu\rho}) - 4 \alpha' \p_z \p_{\overline{z}} \omega \, x^\mu \p_\mu F \notag \\[2pt]
		& \hspace{3cm} + \tfrac{1}{2} \lambda \, \p_{\overline{z}} x^\mu x^\nu \bigl( \p_{\nu} \tau_{\mu} - \p_\mu \tau_\nu \bigr) + \tfrac{1}{2} \, \overline{\lambda} \, \p_{{z}} x^\mu x^\nu \bigl( \p_{\nu} \overline{\tau}_{\mu} - \p_\mu \overline{\tau}_\nu \bigr) \notag \\[2pt]
		& \hspace{7.2cm} - \tfrac{1}{2} \, \bigl( \p_{\overline{z}} \lambda \, \p_\nu \tau_\mu + \p_z \overline{\lambda} \, \p_\nu \overline{\tau}_\mu \bigr) x^\mu x^\nu \Bigr]\,.
\end{align}
Note that we already omitted a term that contains $\p_\mu \p_\nu \tau_\rho$ here.
This $S^{(3)}$ contributes to the path integral via
\be \label{eq:Z(2)cubic}
	\mathcal{Z}^{(2)}_\text{cubic} = \frac{1}{2} \bigl\langle S^{(3)} S^{(3)} \bigr\rangle\,.
\ee
Then, we consider the contributions to $\delta_W \tau_\mu$ ($\delta_W \overline{\tau}_\mu$) from $\mathcal{Z}^{(2)}_\text{cubic}$\,, in front of operators that contain $\lambda$ ($\overline{\lambda}$). These contributions necessarily involve the operator $\lambda \, \p_{\overline{z}} x^\mu x^\nu$ or $\overline{\lambda} \, \p_{{z}} x^\mu x^\nu$ in \eqref{eq:S(3)}. For example, $\mathcal{Z}^{(2)}_\text{cubic}$ contains 
\begin{align} \label{eq:lambdadwx}
	\mathcal{Z}^{(2)}_\text{cubic} & \supset \int d^2 z : \lambda \, \partial_{\overline{z}} x^\mu x^\nu : \p_{[\nu} \tau_{\mu]} \int d^2 w : \partial_w x^\sigma \partial_{\overline{w}} x^\lambda x^\rho : \p_\rho H_{\sigma\lambda} \notag \\
		& \supset \int d^2 z \, d^2 w \frac{1}{|z - w|^2} \, : \lambda (z) \, \p_{\overline{w}} x^\mu (w, \overline{w}) : \p^{}_{[B'} \tau^{}_{A']} \p^{A'} H^{B'}{}_{\mu}\,,
\end{align}
Cutting the integral by an invariant distance $\epsilon$ such that $|z - w| \, e^\omega > \epsilon$ introduces a scale dependence of the metric, $\ln |z - w|_\text{min} = - \omega + \ln \epsilon$\, \cite{Polchinski:1998rq}. Hence, we obtain a contribution to $\delta_W \tau_\mu$ from \eqref{eq:lambdadwx},\,\footnote{One also needs to impose the path integral identity \eqref{eq:addid} to write down the final contribution.} 
\be \label{eq:dtaudH}
	\delta_W \tau_\mu \supset \p^{}_{[B'} \tau^{}_{A']} \p^{A'} H^{B'}{}_{\mu}\,.
\ee
Note that the coefficient in front of $\lambda \, \p_{\overline{z}} x^\mu x^\nu$ in \eqref{eq:S(3)} is proportional to $\p_{[\mu} \tau_{\nu]}$\,, whose covariantization is $D_{[\mu} \tau_{\nu]}$\,. The analogy of this observation also holds for the operator $\overline{\lambda} \, \p_z x^\mu x^\nu$ in \eqref{eq:S(3)}. 
Hence, contributions to $\delta_W \tau_\mu{}^A$ from the operators $\lambda \, \p_{\overline{z}} x^\mu x^\nu$ and $\overline{\lambda} \, \p_{{z}} x^\mu x^\nu$\,, such as the one in \eqref{eq:lambdadwx}, necessarily contain components in $D_{[\mu} \tau_{\nu]}{}^A$\,, and these contributions vanish when $D_{[\mu} \tau_{\nu]}{}^A = 0$ is imposed. Therefore, we conclude that any contribution to $\delta_W \tau_\mu{}^A$ that involves $\p_\mu g_i \p_\nu g_i$ has to vanish when $D_{[\mu} \tau_{\nu]}{}^A = 0$ is imposed. This demonstrates that the equations from vanishing $\beta^\tau_{(AB)}$ and $\beta^\tau_{A'A}$ are solved by the hypersurface orthogonality condition $D_{[\mu} \tau_{\nu]}{}^A = 0$\,.

In the last line of $S^{(3)}$ in \eqref{eq:S(3)}, there are terms that contain $\p_{\overline{z}} \lambda$ and $\p_z \overline{\lambda}$\,. These terms contribute to the dilaton beta-function. For example, $Z^{(2)}_\text{cubic}$ contains the following term:
\begin{align} \label{eq:Zcubiclambdax}
	\mathcal{Z}^{(2)}_\text{cubic} & \supset \int d^2 z : \partial_{\overline{z}} \lambda \, x^\mu x^\nu : \p_\nu \tau_\mu \int d^2 w : \partial_w x^\sigma \partial_{\overline{w}} x^\lambda x^\rho : \p_\rho H_{\sigma\lambda} \notag \\
		& \supset \int d^2 z \, d^2 w \frac{1}{|z - w|^2} \, : \p_{\overline{z}} \lambda (z) \, x^\mu (w, \overline{w}) : \p_{A'} \tau_{B'} \p_\mu H^{A'B'}\,.
\end{align}
Taking a Fourier transform of the path integral identity in \eqref{eq:eq:renormeom2}, we obtain that in position space,
\be \label{eq:gammaF}
	\bigl[ \p_{\overline{z}} \lambda (z) \, x^\mu (w, \overline{w}) \bigr]_\text{r} \propto \gamma R^{(2)} (\delta^\mu_0 + \delta^\mu_1)\,,
\ee
applying which to \eqref{eq:Zcubiclambdax} generates a contribution to $\delta_W F$ that is dependent on the renormalization scheme. We will come back to this issue at the end of this section.

In \S\ref{sec:fevo}, we set the coupling 
\be \label{eq:couplingU}
	U \equiv - 4 \pi g \, \int d^d K \, u (K) \, e^{i K_\mu x^\mu}
\ee
in front of the operator $\lambda \overline{\lambda}$ to zero and then showed that this operator is not generated at linear order.
However, if $D_{[\mu} \tau_{\nu]}{}^A = 0$ were not imposed \emph{a priori}, the operator $\lambda \overline{\lambda}$ would have received nontrivial quantum corrections from the following higher order contribution in $\mathcal{Z}^{(2)}_\text{cubic}$\,:
\begin{align} \label{eq:Zcubic(2)eva}
	\mathcal{Z}^{(2)}_\text{cubic} & \supset \int d^2z : \lambda \, \p_{\overline{z}} x^\mu x^\nu \p_{[\nu} \tau_{\mu]} : \int d^2w : \overline{\lambda} \, \p_{w} x^\rho x^{\sigma} \p_{[\sigma}\! \overline{\tau}_{\rho]} : \notag \\
		& \supset \int d^2 z \, d^2 w \, \frac{1}{|z - w|^2} : \lambda (z) \, \overline{\lambda} (\overline{w}) : \p_{[\Ap} \tau_{\Bp]} \, \p^{[\Bp} \overline{\tau}^{\Ap]}\,.
\end{align}
Note that the covariantization of $\p_{[\Ap} \tau_{\Bp]} \, \p^{[\Bp} \overline{\tau}^{\Ap]}$ is $E^{\mu\nu} E^{\rho\sigma} D_{[\mu} \tau_{\rho]} D_{[\nu} \overline{\tau}_{\sigma]}$\,,
which vanishes identically once $D_{[\mu} \tau_{\nu]}{}^A = 0$ is imposed, as required by the $Z_A$ gauge symmetry. This shows explicitly that the operator $\lambda \overline{\lambda}$ is not generated at loop level, in consistency with the fact that $U = 0$ is protected by the $Z_A$ symmetry.

We already showed that $\beta^\tau_{(AB)} = \beta^\tau_{\Ap A} = 0$ is solved by $D_{[\mu} \tau_{\nu]}{}^A = 0$\,.\,\footnote{It is impossible to have solutions other than $D_{[\mu} \tau_{\nu]}{}^A = 0$ without breaking the $Z_A$ gauge symmetry. We have shown that, without the $Z_A$ symmetry, one would have to turn on the $\lambda \overline{\lambda}$ operator in the Lagrangian to have enough counterterms to absorb all the divergences, which generates a deformation towards relativistic string theory. This is beyond the scope of this paper.} In the following, we will once for all set $D_{[\mu} \tau_{\nu]}{}^A = 0$\,, which significantly simplifies the remaining beta-functions. Under this condition, we collect all the ingredients that contain two covariant derivatives as follows: 
\be
	R_{\mu\nu} = R^\rho{}_{\mu\rho\nu}\,,
		\quad
	\nabla_\rho \CH_{\sigma\mu\nu}\,, 
		\quad
	\nabla_\mu \nabla_\nu \Phi\,,
		\quad
	\CH_{\mu\rho\kappa} \CH_{\nu\sigma\lambda}\,,
		\quad
	\nabla_\mu \Phi \, \CH_{\nu\rho\sigma}\,,
		\quad
	\nabla_\mu \Phi \, \nabla_\nu \Phi\,.
\ee 
Following the standard treatment in \cite{Polchinski:1998rq}, by evaluating the associated terms in \eqref{eq:Z(2)cubic} and then covariantizing the results using the Riemann tensor $R^\rho{}_{\sigma\mu\nu}$ and the covariant derivative $\nabla_\mu$\,, we find that \eqref{eq:betafunction} becomes
\begin{subequations} \label{eq:fullbetafunctions}
\begin{align}
	\beta^H_{\Ap \Bp} \! & = \alpha' P_{\Ap\Bp} + O({\alpha'}^2) \,, \\[8pt]
	\beta^B_{\Ap \Bp} \! & = \alpha' Q_{\Ap\Bp} + O({\alpha'}^2) \,, \\[5pt]
	\beta^\Theta_{A\Ap} \! & = \frac{\alpha'}{2} \, \bigl( P_{A\Ap} + \epsilon_A{}^B Q_{B\Ap} ) + O({\alpha'}^2) \,, \\[5pt]
	\beta^\Theta \! & = - \frac{\alpha'}{4} \bigl( \eta^{AB} P_{AB} - \epsilon^{AB} Q_{AB}  \bigr) + O({\alpha'}^2)\,, \\[5pt]
	\beta^F \! & = \frac{d - \! 26}{6} \! - \alpha' \bigl( \nabla^{}_{\!\Ap} \nabla^\Ap \Phi - \nabla^\Ap \Phi \nabla_{\!\Ap} \Phi + \tfrac{1}{4} R_{\Ap}{}^{\Ap} \!\! - \tfrac{1}{48} \CH_{\Ap\Bp\Cp} \CH^{\Ap\Bp\Cp} \bigr) \notag \\
	 	& \hspace{1.65cm} + O({\alpha'}^2)\,, \! \label{eq:betaF1}
\end{align}
\end{subequations}
where
\begin{subequations}
\begin{align}
	P_{\mu\nu} & \equiv R_{\mu\nu} + 2 \nabla_\mu \nabla_\nu \Phi - \tfrac{1}{4} \mathcal{H}_{\mu\Ap\Bp} \mathcal{H}_{\nu}{}^{\Ap\Bp} \,, \\[3pt]
	Q_{\mu\nu} & \equiv - \tfrac{1}{2} \nabla^\Ap \mathcal{H}_{\Ap\mu\nu} + \nabla^\Ap \Phi \, \mathcal{H}_{\Ap\mu\nu} \,.
\end{align}
\end{subequations}

We point out one shortcut in the derivation of the beta-functions \eqref{eq:fullbetafunctions}. Earlier in \eqref{eq:gammaF}, we found that there are renormalization scheme dependent contributions to $\beta^F$ that involve $\tau_\mu{}^A$\,. Moreover, the evaluation of the coefficient of the last term $\CH_{\Ap\Bp\Cp} \CH^{\Ap\Bp\Cp}$ in \eqref{eq:betaF1} requires applying the renormalization scheme with the choice $\gamma = -1$ made in \eqref{eq:betafunction}. Regardless of these subtleties, there is an elegant way of fixing the coefficient of $\CH_{\Ap\Bp\Cp} \CH^{\Ap\Bp\Cp}$ and demonstrating that there are no extra contributions from $\tau_\mu{}^A$ to $\beta^F$ at the same time. On a locally flat worldsheet, the dilaton is inaccessible, and conformal invariance only requires $\beta^H_{\Ap\Bp} = \beta^B{}_{\Ap\Bp} = \beta^\Theta_{A\Ap} = \beta^\Theta = 0$\,, in addition to the geometric constraints that we already enforced to preserve the $Z_A$ symmetry. These conditions are also sufficient for maintaining conformal invariance on a curved worldsheet. Therefore, $\beta^F = 0$ has to be dependent of other vanishing beta-functions -- the central charge is the only additional contribution from the curvature of the worldsheet, which we set to zero by requiring $d=26$\,.\,\footnote{As in relativistic string theory, there also exist other solutions with, for example, $\Phi (x) = v_\mu x^\mu$\,, with $\alpha' v^\Ap v_\Ap = (26 - d)/6$\,.} 
The $\Phi$-dependent terms in \eqref{eq:betaF1} determine the consistency equation to be \cite{Deligne:1999qp}
\be \label{eq:consisrelation}
	\nabla^\Bp \beta^H_{\Ap\Bp} = - 2 \nabla_\Ap \beta^F + \frac{1}{2} \, \beta^B_{\Bp\Cp} \CH_\Ap{}^{\Bp\Cp} + 2 \beta^H_{\Ap\Bp} \nabla^\Bp \Phi\,.
\ee
Applying the Bianchi identities,
\begin{subequations}
\begin{align}
	\nabla^\Bp R_{\Ap\Bp} & = \tfrac{1}{2} \nabla_\Ap R_\Bp{}^\Bp\,, \\[2pt]
	\nabla^\Bp \bigl( \CH_{\Bp \Cp \Dp } \CH_\Ap{}^{\Cp \Dp } \bigr) & = \nabla^\Bp \CH_{\Bp \Cp \Dp } \CH_\Ap{}^{\Cp \Dp } + \tfrac{1}{6} \nabla_\Ap \bigl( \CH_{\Bp\Cp \Dp } \CH^{\Bp\Cp \Dp } \bigr)\,, 
\end{align}
\end{subequations}
we find that the beta-functions in \eqref{eq:fullbetafunctions} satisfy \eqref{eq:consisrelation}. This procedure not only fixes the coefficient in front of $\mathcal{H}_{\Ap\Bp\Cp} \mathcal{H}^{\Ap\Bp\Cp}$ in \eqref{eq:betaF1} but also excludes any additional contribution from $\tau_\mu{}^A$ in \eqref{eq:betaF1}.

In summary, we collect the independent classical equations of motion for the background string Newton-Cartan geometry, $B$-field and dilaton:
\begin{subequations} \label{eq:resultbf}
\begin{align}
	\epsilon^{(A}{}^{}_C \, \tau^{}_{\,[\mu}{}^{B)} \, \p^{}_{\nu\phantom{]}\!} \tau^{}_{\rho]}{}^C & = 0\,, \\[2pt]
	P_{\Ap\Bp} = Q_{\Ap\Bp} = P_{A\Ap} + \epsilon_A{}^B Q_{B\Ap} = \eta^{AB} P_{AB} - \epsilon^{AB} Q_{AB} & = 0\,, \\[2pt]
	\nabla_\Ap \nabla^\Ap \Phi - \nabla^\Ap \Phi \nabla_\Ap \Phi + \tfrac{1}{4} R_\Ap{}^\Ap - \tfrac{1}{48} \CH_{\Ap\Bp\Cp} \CH^{\Ap\Bp\Cp} & = \frac{d-26}{6 \alpha'}\,.
\end{align}
\end{subequations}

\section{Conclusions and Discussions} \label{sec:conclusions}

In this paper, we studied the worldsheet quantum consistency of the classically Weyl invariant sigma model of nonrelativistic string theory. Requiring that the worldsheet theory is Weyl invariant at quantum level, we derive at one-loop the background equations of motion together with the hypersurface orthogonality condition $D_{[\mu} \tau_{\nu]}{}^A = 0$\,,\,\footnote{This condition is required by the $Z_A$ gauge symmetry.} in string Newton-Cartan gravity coupled to the Kalb-Ramond and dilaton field. Such equations of motion determine the backgrounds on which nonrelativistic string theory can be consistently defined. In a companion paper \cite{stringyNClimit}, it is shown that the same set of equations of motion can be derived as a subtle limit of relativistic beta-functions.

It is intriguing to note that the combinations of the string Newton-Cartan geometry, Kalb-Ramond and dilaton field in \eqref{eq:defineThetaF} that have nontrivial beta-functions are reminiscent of ingredients in double field theory formalism. For example, it has been shown in \cite{nrDFT, Morand:2017fnv, Park:2016sbw} that nonrelativistic string theory in flat space can be embedded in the double field theory formalism. A more thorough study of these connections may in turn deepen the understanding of T-dualities in nonrelativistic string theory, which was recently formulated using a first principles method in \cite{Bergshoeff:2018yvt}. 

Another important question to ask is whether there exists a target space effective action whose variation gives the equations of motion in string Newton-Cartan gravity. The action principle is usually unavailable for (string) Newton-Cartan gravity theories. Nevertheless, exceptions exist. For examples, actions for extensions of Newton-Cartan gravity\,\footnote{In Newton-Cartan gravity, there is only one ``longitudinal" dimension, i.e. the temporal direction. In contrast, in string Newton-Cartan gravity, there are two longitudinal dimensions.} have been constructed in three dimensions in \cite{Papageorgiou:2009zc, Bergshoeff:2016lwr, Hartong:2016yrf, Ozdemir:2019orp}, and in general dimensions for a novel nonrelativistic algebra in \cite{
Hansen:2018ofj}.\,\footnote{A systematic method of generating actions for Newton-Cartan-type of theories has been explored in \cite{Bergshoeff:2019ctr}.} More recently, an action for the so-called four-dimensional extended string Newton-Cartan gravity was constructed in \cite{Bergshoeff:2018vfn}, paying the price of introducing an extra vector field associated with a central extension in the string Newton-Cartan algebra. One possibility is that this extra vector is responsible for some extra anyonic strings in the spectrum. A similar mechanism was realized in relativistic string theory in three spacetime dimensions \cite{Mezincescu:2010yp}: the spectrum contains anyonic particles that are absent in higher dimensions. 

The string Newton-Cartan geometry arises from the condensate of zero winding intermediate states that meditate instantaneous gravitational forces among on-shell winding closed strings.
It would be interesting to explore the possibility of introducing nonzero winding numbers in the analysis in \S\ref{sec:fevo}, which may allow us to probe more general geometries and shed light on spacetime (string) field theory.  

We close the paper with a few more open questions. First, it would be useful to develop a reliable background field method 
which is helpful for extracting contributions from higher loop corrections to the beta-functions, and hence higher derivative corrections to string Newton-Cartan gravity. Second, it would be interesting to derive the beta-functions in the presence of either longitudinal 
or transverse 
D-branes, which should add nonrelativistic twists to the usual Dirac-Born-Infeld effective field theory on D-branes. Finally, it is also natural to generalize this framework to incorporate worldsheet or spacetime supersymmetries and look for solutions with a non-trivial horizon, which may bare interesting applications to holography.
 
\acknowledgments

We would like to thank Eric Bergshoeff, Lorenzo Di Pietro, Zachary Fisher, Laurent Freidel, Kevin T. Grosvenor, Troels Harmark, Charles Melby-Thompson, Tadashi Okazaki, Jan Rosseel and Ceyda \c{S}im\c{s}ek for useful discussions. ZY would like to thank Petr Ho\v{r}ava and Berkeley Center for Theoretical Physics for hospitability. This research was supported in part by Perimeter Institute for Theoretical Physics. Research at Perimeter Institute is supported by the Government of Canada through the Department of Innovation, Science and Economic Development and by the Province of Ontario through the Ministry of Research, Innovation and Science. The research of JO is supported in part by Kwanjeong Educational Foundation and by the Visiting Graduate Fellowship Program at the Perimeter Institute for Theoretical Physics.

\newpage

\bibliographystyle{JHEP}
\bibliography{nrstbf}

\providecommand{\href}[2]{#2}\begingroup\raggedright\begin{thebibliography}{10}

\bibitem{Gomis:2000bd}
J.~Gomis and H.~Ooguri, \emph{{Nonrelativistic closed string theory}},
  \href{https://doi.org/10.1063/1.1372697}{\emph{J. Math. Phys.} {\bfseries 42}
  (2001) 3127} [\href{https://arxiv.org/abs/hep-th/0009181}{{\ttfamily
  hep-th/0009181}}].

\bibitem{Klebanov:2000pp}
I.~R. Klebanov and J.~M. Maldacena, \emph{{(1+1)-dimensional NCOS and its U(N)
  gauge theory dual}}, \href{https://doi.org/10.4310/ATMP.2000.v4.n2.a3,
  10.1142/S0217751X0100400X, 10.1142/S0217751X01004001}{\emph{Int. J. Mod.
  Phys.} {\bfseries A16} (2001) 922}
  [\href{https://arxiv.org/abs/hep-th/0006085}{{\ttfamily hep-th/0006085}}].

\bibitem{Danielsson:2000gi}
U.~H. Danielsson, A.~Guijosa and M.~Kruczenski, \emph{{IIA/B, wound and
  wrapped}}, \href{https://doi.org/10.1088/1126-6708/2000/10/020}{\emph{JHEP}
  {\bfseries 10} (2000) 020}
  [\href{https://arxiv.org/abs/hep-th/0009182}{{\ttfamily hep-th/0009182}}].

\bibitem{Gomis:2005pg}
J.~Gomis, J.~Gomis and K.~Kamimura, \emph{{Non-relativistic superstrings: a new
  soluble sector of $AdS_5 \times S^5$}},
  \href{https://doi.org/10.1088/1126-6708/2005/12/024}{\emph{JHEP} {\bfseries
  12} (2005) 024} [\href{https://arxiv.org/abs/hep-th/0507036}{{\ttfamily
  hep-th/0507036}}].

\bibitem{nrGalilei}
J.~Brugues, T.~Curtright, J.~Gomis and L.~Mezincescu, \emph{{Non-relativistic
  strings and branes as non-linear realizations of Galilei groups}},
  \href{https://doi.org/10.1016/j.physletb.2004.05.024}{\emph{Phys. Lett.}
  {\bfseries B594} (2004) 227}
  [\href{https://arxiv.org/abs/hep-th/0404175}{{\ttfamily hep-th/0404175}}].

\bibitem{stringyNC}
R.~Andringa, E.~Bergshoeff, J.~Gomis and M.~de~Roo, \emph{{`Stringy'
  Newton-Cartan gravity}},
  \href{https://doi.org/10.1088/0264-9381/29/23/235020}{\emph{Class. Quant.
  Grav.} {\bfseries 29} (2012) 235020}
  [\href{https://arxiv.org/abs/arXiv:1206.5176}{{\ttfamily arXiv:1206.5176}}].

\bibitem{NCbranes}
J.~Brugues, J.~Gomis and K.~Kamimura, \emph{{Newton-Hooke algebras,
  non-relativistic branes and generalized pp-wave metrics}},
  \href{https://doi.org/10.1103/PhysRevD.73.085011}{\emph{Phys. Rev.}
  {\bfseries D73} (2006) 085011}
  [\href{https://arxiv.org/abs/hep-th/0603023}{{\ttfamily hep-th/0603023}}].

\bibitem{stringyNClimit}
E.~A. Bergshoeff, J.~Gomis, J.~Rosseel, C.~\c{S}im\c{s}ek and Z.~Yan,
  \emph{{String Theory and String Newton-Cartan Geometry}},
  \href{https://arxiv.org/abs/arXiv:1907.10668}{{\ttfamily arXiv:1907.10668}}.

\bibitem{Bergshoeff:2018yvt}
E.~Bergshoeff, J.~Gomis and Z.~Yan, \emph{{Nonrelativistic string theory and
  T-duality}}, \href{https://doi.org/10.1007/JHEP11(2018)133}{\emph{JHEP}
  {\bfseries 11} (2018) 133}
  [\href{https://arxiv.org/abs/arXiv:1806.06071}{{\ttfamily
  arXiv:1806.06071}}].

\bibitem{Batlle:2016iel}
C.~Batlle, J.~Gomis and D.~Not, \emph{{Extended Galilean symmetries of
  non-relativistic strings}},
  \href{https://doi.org/10.1007/JHEP02(2017)049}{\emph{JHEP} {\bfseries 02}
  (2017) 049} [\href{https://arxiv.org/abs/arXiv:1611.00026}{{\ttfamily
  arXiv:1611.00026}}].

\bibitem{Gomis:2016zur}
J.~Gomis and P.~K. Townsend, \emph{{The Galilean superstring}},
  \href{https://doi.org/10.1007/JHEP02(2017)105}{\emph{JHEP} {\bfseries 02}
  (2017) 105} [\href{https://arxiv.org/abs/arXiv:1612.02759}{{\ttfamily
  arXiv:1612.02759}}].

\bibitem{Batlle:2017cfa}
C.~Batlle, J.~Gomis, L.~Mezincescu and P.~K. Townsend, \emph{{Tachyons in the
  Galilean limit}}, \href{https://doi.org/10.1007/JHEP04(2017)120}{\emph{JHEP}
  {\bfseries 04} (2017) 120}
  [\href{https://arxiv.org/abs/arXiv:1702.04792}{{\ttfamily
  arXiv:1702.04792}}].

\bibitem{HHO}
T.~Harmark, J.~Hartong and N.~A. Obers, \emph{{Nonrelativistic strings and
  limits of the AdS/CFT correspondence}},
  \href{https://doi.org/10.1103/PhysRevD.96.086019}{\emph{Phys. Rev.}
  {\bfseries D96} (2017) 086019}
  [\href{https://arxiv.org/abs/arXiv:1705.03535}{{\ttfamily
  arXiv:1705.03535}}].

\bibitem{Kluson}
J.~Kluso\v{n}, \emph{{Remark about non-relativistic string in Newton-Cartan
  background and null reduction}},
  \href{https://arxiv.org/abs/arXiv:1803.07336}{{\ttfamily arXiv:1803.07336}}.

\bibitem{Kluson:2018grx}
J.~Kluso\v{n}, \emph{{Nonrelativistic string theory sigma model and its
  canonical formulation}},
  \href{https://doi.org/10.1140/epjc/s10052-019-6623-9}{\emph{Eur. Phys. J.}
  {\bfseries C79} (2019) 108}
  [\href{https://arxiv.org/abs/arXiv:1809.10411}{{\ttfamily
  arXiv:1809.10411}}].

\bibitem{Harmark:2018cdl}
T.~Harmark, J.~Hartong, L.~Menculini, N.~A. Obers and Z.~Yan, \emph{{Strings
  with non-relativistic conformal symmetry and limits of the AdS/CFT
  correspondence}}, \href{https://doi.org/10.1007/JHEP11(2018)190}{\emph{JHEP}
  {\bfseries 11} (2018) 190}
  [\href{https://arxiv.org/abs/arXiv:1810.05560}{{\ttfamily
  arXiv:1810.05560}}].

\bibitem{Kluson:2018vfd}
J.~Kluso\v{n}, \emph{{Note about T-duality of non-relativistic string}},
  \href{https://arxiv.org/abs/arXiv:1811.12658}{{\ttfamily arXiv:1811.12658}}.

\bibitem{Kluson:2019ifd}
J.~Kluso\v{n}, \emph{{$(m,n)$-string and D1-brane in stringy Newton-Cartan
  background}},  \href{https://arxiv.org/abs/arXiv:1901.11292}{{\ttfamily
  arXiv:1901.11292}}.

\bibitem{Roychowdhury:2019qmp}
D.~Roychowdhury, \emph{{Probing tachyon kinks in Newton-Cartan background}},
  \href{https://arxiv.org/abs/arXiv:1903.05890}{{\ttfamily arXiv:1903.05890}}.

\bibitem{Roychowdhury:2019vzh}
D.~Roychowdhury, \emph{{On integrability in nonrelativistic string theory}},
  \href{https://arxiv.org/abs/arXiv:1904.06485}{{\ttfamily arXiv:1904.06485}}.

\bibitem{Harmark:2019upf}
T.~Harmark, J.~Hartong, L.~Menculini, N.~A. Obers and G.~Oling, \emph{{Relating
  non-relativistic string theories}},
  \href{https://arxiv.org/abs/arXiv:1907.01663}{{\ttfamily arXiv:1907.01663}}.

\bibitem{Gallegos:2019icg}
A.~D. Gallegos, U.~Gursoy and N.~Zinnato, \emph{{Torsional Newton Cartan
  gravity from non-relativistic strings}},
  \href{https://arxiv.org/abs/arXiv:1906.01607}{{\ttfamily arXiv:1906.01607}}.

\bibitem{Polchinski:1998rq}
J.~Polchinski, \emph{{String theory. Vol. 1: An introduction to the bosonic
  string}}, Cambridge Monographs on Mathematical Physics. Cambridge University
  Press, 2007,
  \href{https://doi.org/10.1017/CBO9780511816079}{10.1017/CBO9780511816079}.

\bibitem{Danielsson:2000mu}
U.~H. Danielsson, A.~Guijosa and M.~Kruczenski, \emph{{Newtonian gravitons and
  D-brane collective coordinates in wound string theory}},
  \href{https://doi.org/10.1088/1126-6708/2001/03/041}{\emph{JHEP} {\bfseries
  03} (2001) 041} [\href{https://arxiv.org/abs/hep-th/0012183}{{\ttfamily
  hep-th/0012183}}].

\bibitem{Bergshoeff:2018vfn}
E.~A. Bergshoeff, K.~T. Grosvenor, C.~\c{S}im\c{s}ek and Z.~Yan, \emph{{An
  action for extended string Newton-Cartan gravity}},
  \href{https://doi.org/10.1007/JHEP01(2019)178}{\emph{JHEP} {\bfseries 01}
  (2019) 178} [\href{https://arxiv.org/abs/arXiv:1810.09387}{{\ttfamily
  arXiv:1810.09387}}].

\bibitem{Deligne:1999qp}
P.~Deligne, P.~Etingof, D.~S. Freed, L.~C. Jeffrey, D.~Kazhdan, J.~W. Morgan
  et~al., eds., \emph{{Quantum fields and strings: A course for mathematicians.
  Vol. 1, 2}}. American Mathematical Society, 1999.

\bibitem{nrDFT}
S.~M. Ko, C.~Melby-Thompson, R.~Meyer and J.-H. Park, \emph{Dynamics of
  perturbations in double field theory \& non-relativistic string theory},
  \href{https://doi.org/10.1007/JHEP12(2015)144}{\emph{JHEP} {\bfseries 12}
  (2015) 144} [\href{https://arxiv.org/abs/arXiv:1508.01121}{{\ttfamily
  arXiv:1508.01121}}].

\bibitem{Morand:2017fnv}
K.~Morand and J.-H. Park, \emph{{Classification of non-Riemannian
  doubled-yet-gauged spacetime}},
  \href{https://doi.org/10.1140/epjc/s10052-017-5257-z,
  10.1140/epjc/s10052-018-6394-8}{\emph{Eur. Phys. J.} {\bfseries C77} (2017)
  685} [\href{https://arxiv.org/abs/arXiv:1707.03713}{{\ttfamily
  arXiv:1707.03713}}].

\bibitem{Park:2016sbw}
J.-H. Park, \emph{{Green-Schwarz superstring on doubled-yet-gauged spacetime}},
  \href{https://doi.org/10.1007/JHEP11(2016)005}{\emph{JHEP} {\bfseries 11}
  (2016) 005} [\href{https://arxiv.org/abs/arXiv:1609.04265}{{\ttfamily
  arXiv:1609.04265}}].

\bibitem{Papageorgiou:2009zc}
G.~Papageorgiou and B.~J. Schroers, \emph{{A Chern-Simons approach to Galilean
  quantum gravity in 2+1 dimensions}},
  \href{https://doi.org/10.1088/1126-6708/2009/11/009}{\emph{JHEP} {\bfseries
  11} (2009) 009} [\href{https://arxiv.org/abs/arXiv:0907.2880}{{\ttfamily
  arXiv:0907.2880}}].

\bibitem{Bergshoeff:2016lwr}
E.~A. Bergshoeff and J.~Rosseel, \emph{{Three-dimensional extended Bargmann
  supergravity}},
  \href{https://doi.org/10.1103/PhysRevLett.116.251601}{\emph{Phys. Rev. Lett.}
  {\bfseries 116} (2016) 251601}
  [\href{https://arxiv.org/abs/arXiv:1604.08042}{{\ttfamily
  arXiv:1604.08042}}].

\bibitem{Hartong:2016yrf}
J.~Hartong, Y.~Lei and N.~A. Obers, \emph{{Nonrelativistic Chern-Simons
  theories and three-dimensional Ho\v{r}ava-Lifshitz gravity}},
  \href{https://doi.org/10.1103/PhysRevD.94.065027}{\emph{Phys. Rev.}
  {\bfseries D94} (2016) 065027}
  [\href{https://arxiv.org/abs/arXiv:1604.08054}{{\ttfamily
  arXiv:1604.08054}}].

\bibitem{Ozdemir:2019orp}
N.~Ozdemir, M.~Ozkan, O.~Tunca and U.~Zorba, \emph{{Three-Dimensional Extended
  Newtonian (Super)Gravity}},
  \href{https://doi.org/10.1007/JHEP05(2019)130}{\emph{JHEP} {\bfseries 05}
  (2019) 130} [\href{https://arxiv.org/abs/arXiv:1903.09377}{{\ttfamily
  arXiv:1903.09377}}].

\bibitem{Hansen:2018ofj}
D.~Hansen, J.~Hartong and N.~A. Obers, \emph{{An action principle for Newtonian
  gravity}}, \href{https://doi.org/10.1103/PhysRevLett.122.061106}{\emph{Phys.
  Rev. Lett.} {\bfseries 122} (2019) 061106}
  [\href{https://arxiv.org/abs/arXiv:1807.04765}{{\ttfamily
  arXiv:1807.04765}}].

\bibitem{Bergshoeff:2019ctr}
E.~Bergshoeff, J.~M. Izquierdo, T.~Ortin and L.~Romano, \emph{{Lie algebra
  expansions and actions for non-relativistic gravity}},
  \href{https://arxiv.org/abs/arXiv:1904.08304}{{\ttfamily arXiv:1904.08304}}.

\bibitem{Mezincescu:2010yp}
L.~Mezincescu and P.~K. Townsend, \emph{{Anyons from strings}},
  \href{https://doi.org/10.1103/PhysRevLett.105.191601}{\emph{Phys. Rev. Lett.}
  {\bfseries 105} (2010) 191601}
  [\href{https://arxiv.org/abs/arXiv:1008.2334}{{\ttfamily arXiv:1008.2334}}].

\end{thebibliography}\endgroup

\end{document}